\documentclass[aip,jcp,preprint,notitlepage,superscriptaddress,amsmath,amssymb]{revtex4-2}

\usepackage{graphicx}
\usepackage{graphics}
\usepackage{slashbox}

\usepackage[utf8x]{inputenc}
\usepackage{amsmath,graphicx}
\usepackage{amsfonts}
\usepackage{amssymb}
\usepackage{color}
\usepackage{multirow}
\usepackage[version=3]{mhchem} 
\usepackage{xcolor}

\usepackage{dcolumn}
\usepackage{bm}
\usepackage{natbib}
\usepackage{tabulary}
\usepackage{textcomp}
\usepackage{csquotes}
\newcommand{\stano}[1]{{\color{black}{#1}}}
\newcommand{\stanoo}[1]{{\color{black}{#1}}}
\newcommand{\shota}[1]{{\color{black}{#1}}}
\newcommand{\jozef}[1]{{\color{black}{#1}}}
\def\lbl#1{\label{#1}}

\def\be{\begin{equation}}
\def\ee{\end{equation}}
\def\bee{\begin{equation}}
\def\eee{\end{equation}}
\def\bea{\begin{eqnarray}}
\def\eea{\end{eqnarray}}
\def\bra  #1 {\langle {#1} \vert}
\def\ket #1 {\vert {#1} \rangle}
\def\half{\textstyle{1\over 2}}
\def\quat{\textstyle{1\over 4}}
\def\eighth{\textstyle{1\over 8}}
\def\eqaux{\buildrel {\tiny\mathrm{cabs}} \over{=}}
\def\eqsa{\buildrel {\tiny\mathrm{s.a.}} \over{=}}

\def\bea{\begin{eqnarray}}
\def\eea{\end{eqnarray}}
\def\cQ{\mathcal Q}
\def\cD{\mathcal D}
\def\cE{\mathcal E}
\def\cW{\mathcal W}
\def\cG{\mathcal G}
\def\cI{\mathcal I}
\def\cJ{\mathcal J}
\def\cK{\mathcal K}
\def\cL{\mathcal L}
\def\cM{\mathcal M}

\def\cR {{\mathcal R}}
\def\cV {{\mathcal V}}
\def\cW {{\mathcal W}}
\def\cG {{\mathcal G}}
\def\cI {{\mathcal I}}

\def\cB {{\mathcal B}}

\def\cP {{\mathcal P}}

\def\cZ {{\mathcal Z}}

\def\cX {{\mathcal X}}

\def\cU{\mathcal U}
\def\cY{\mathcal Y}

\def\cG{\mathcal G}
\def\mF{\mathcal F}
\def\tableline{\noalign{\smallskip\hrule\smallskip} }   
\def\nos{\noalign{\smallskip}}                          

\begin{document}
\title{An Alternative Explicitly Correlated Approach for Describing Strongly Correlated Systems within a Single-Determinant Framework}
\author{Stanislav Ked\v{z}uch}
\email{stanislav.kedzuch@riken.jp}
\affiliation{RIKEN Center for Computational Science, RIKEN, Minatojima-minami 7-1-26, 650-0047, Kobe, Japan}
\author{Shota Tsuru}
\affiliation{RIKEN Center for Computational Science, RIKEN, Minatojima-minami 7-1-26, 650-0047, Kobe, Japan}
\author{Takahito Nakajima}
\affiliation{RIKEN Center for Computational Science, RIKEN, Minatojima-minami 7-1-26, 650-0047, Kobe, Japan}
\author{Jozef Noga} 
\affiliation{Department of Inorganic Chemistry, Faculty of Natural Sciences, Comenius University, SK-84215 Bratislava}
\begin{abstract}
\stano{We introduce an alternative explicitly correlated single reference electronic structure approach designed to describe strongly correlated systems within a single-determinant framework, capable of providing a balanced, robust, and accurate description of both strong and weak electron correlations by integrating the singlet-paired coupled-cluster singles and doubles (CCSD0) methodology with explicitly correlated F12 theory.}
While the singlet (opposite-spin) interactions are captured via standard explicitly correlated terms, the crucial 
\stano{triplet (same-spin) correlation omitted} during the CCSD0 regularization is recovered through a newly formulated explicitly correlated pseudo-excitation operator F12Pi which satisfies the Kato's cusp condition and uniquely integrates a full virtual space to explicitly retrieve the missing triplet-paired electron correlations based on the interelectronic distance $r_{12}$.
We \stanoo{will} 
show that CCSD0 [J. Chem. Theory Comput.  $\mathbf{11}$, 3171 (2015)] \stanoo{is structurally related to the complex molecular orbital (CMO) framework, an orbital theory that addresses the \stanoo{stability} dilemma by yielding an energy lower than that of restricted Hartree–Fock (RHF) in the presence of imaginary spin-singlet instabilities, 
\stanoo{while the symmetry dilemma is resolved via projection} [J. Chem. Phys. $\mathbf{69}$, 4535 (1978)].}  
\stanoo{Prompted by this structural correspondence, we introduce the explicitly correlated "CCSD0-F12Pi" and perturbative triples "CCSD0(T)-F12Pi" methodologies. The performance of these new single-reference frameworks has been 
evaluated for the challenging bond-breaking processes of $\mathrm{N_2}$ and $\mathrm{CO_2}$, as well as benchmarked against Mukherjee’s multireference coupled-cluster method (Mk-CC) for the strongly correlated regimes of linear singlet methylene ($\mathrm{CH_2}$) and cyclobutadiene ($\mathrm{C_4H_4}$)}. 
Our results show that the new framework systematically avoids unphysical artifacts, humps, and oscillations along the potential energy surfaces while accelerating convergence toward the complete basis set limit, offering a robust single-reference alternative to expensive multireference methodologies for closed-shell molecules.
\end{abstract}
\maketitle
\section{Introduction}

Concerning the pervasiveness of strong correlation such as bond cleavage, transition states as well as systems containing transition metals, 
\stano{balanced description of both strong and weak correlation effects is of central significance in quantum chemistry. A compelling strategy to achieve this balance is to extend highly accurate single reference approaches, such as the coupled-cluster (CC) theory \cite{firstCC,firstCC-2,firstCC-3,pople1978,bartlett_purvis1978,ccsd,ccsdtpert1,ccsdtpert2,ccsdtparent1,Bartlett2007,Shavitt_Bartlett_2009}},
into a multireference framework \jozef{(MRCC)\cite{MkMRCC-1,MkMRCC-2,MkMRCC-3,CC_book2010,MRCC2012,maitra_mukherjee,ss-MRCC,evangelista,MRCC2025} .} 
\jozef{Among the various models of MRCC, especially Mukherjee’s 
method (Mk-MRCC) \cite{MkMRCC-1,MkMRCC-2,MkMRCC-3,maitra_mukherjee} has been frequently advocated due to its  size-extensivity and "relatively simple" 
computing implementation \cite{evangelista} .}
\jozef{Nevertheless, MRCC remains challenging due to technical problems including those generally inherent to MR theories
 such as choice of the active space and exponential scaling of the computational cost. Bridging the gap between 
single reference CC and multireference methods which can handle larger active spaces was proposed by Li in his
 ”block-correlated coupled-cluster” (BCCC) theory \cite{BCCC2004}, later also combined with the general valence bond (GVB) 
method \cite{GVB} resulting in ”GVB-BCCC” \cite{GVB-BCCC2020,GVB-BCCC2022,GVB-BCCC2024} .}

\jozef{Having in mind the obstacles related to developing MR theories,
single-reference electronic structure theories attempting to treat
strong correlation have also been proposed}\cite{tsuchimochi_PHF,tsuchimochi2014,tsuchimochi2019,bulik,gomez2016-ccsd0open,gomez2016-FSigCCSD,qiu2016,gomez2017,qiu2017,qiu2018,song2022,DC2013,DC2014,DC2019_1,DC2019_2,DC2021,pCC-1,pCC-2,FCCR2018,FCCR2020,FCCR2024,qvcc-1,qvcc-2,qvcc-3,qvcc-4,qvcc-5}. 
Some
of the proposed theories are based on the idea, which dates back to 1955 when a pioneering work by L\"{o}wdin\cite{lowdin1955} was published, of breaking spin-symmetry of the wavefunction and projection to an eigenstate of the $\mathbf{S}^2$ 
operator
or just spin adaptation. For example, Tsuchimochi formulated a "variation-after-projection" algorithm of the spin-projected Hartree--Fock (HF) theory\cite{tsuchimochi_PHF} and developed a second-order perturbation theory referenced there.\cite{tsuchimochi2014,tsuchimochi2019} Scuseria proposed various kinds of the 
CC
theory where the cluster operator has some spin-adapted form\cite{bulik,gomez2016-ccsd0open,gomez2016-FSigCCSD,qiu2016,gomez2017} as well as the CC theory where the CC amplitude are optimized under the spin projection.\cite{qiu2017,qiu2018,song2022} 
Among the proposed single-reference theories, there are also methods 
where strong correlation is added by subtracting something from the ans\"{a}tze or equations.
For example, Kats made the normal CC singles and doubles (CCSD)\cite{ccsd} capable of describing dissociation curves by eliminating 
exchange integrals unphysical in 
dissociating
molecules from
several diagrams 
of
the $t_2$ amplitude equations and named the method "distinguishable cluster (DC) approximation".\cite{DC2013,DC2014,DC2019_1,DC2019_2,DC2021} 
Scuseria added strong correlation to the CC doubles (CCD) by subtracting all configurations other than the pair excitations, naming the method pair coupled cluster doubles (pCCD).\cite{pCC-1,pCC-2} 
There is also a claim that the failure of the normal single-reference CC in describing strong correlation is largely due to its nonvariational formalism, and Knowles formulated the quasi-variational coupled-cluster (QVCC) theory to remedy the failure.\cite{qvcc-1,qvcc-2,qvcc-3,qvcc-4,qvcc-5}
All these single-reference methods achieved success in describing potential energy curves of some cases especially dissociating molecules. However, it is not clear whether these single-reference methods also describe strong correlation in cases other than bond cleavage. It should 
also be
emphasized that the qualitative successes written above do not necessarily mean that these methods can quantitatively predict chemical properties.

\jozef{Since the} strong correlation is related to the HF instability, it may help analyzing the capabilities of the single-reference electronic structure theories to revisit the discussion about the HF stability condition. Following some cases where the determinant at the energy minimum does not satisfy the symmetries imposed by the Hamiltonians
and considering early mathematical works on the correlation problems,\cite{lowdin1955,lowdin1959}
L\"{o}wdin pointed out that the HF Slater determinant is generally in a dilemma between the two demands that it should have a correct symmetry property and be stationary (symmetry dilemma).\cite{lykos} 
When it comes to cases of closed-shell covalently bonded systems, although Nesbet said that it was unlikely in the same discussion,\cite{lykos} existence of an unrestricted Hartree--Fock (UHF) solution whose energy is lower than the restricted Hartree--Fock (RHF) solution with a correct symmetric property remained an open question. Since then, some researchers worked on formulating the HF stability conditions and more general orbital theories.\cite{thouless,cizek_paldus1967,ostlund,seeger_pople1977,hirao-closed,hirao-open,fukutome,yamada-hirata} 
Prior to the discussion involving L\"{o}wdin and Nesbet,
Thouless derived the HF stability conditions in matrix forms in the second-quantization formalism\cite{thouless}. \v{C}\'{i}\v{z}ek and Paldus re-derived the HF stability conditions from the quadratic expansion of the HF energy and classified the HF instabilities to the spin singlet and triplet instabilities.\cite{cizek_paldus1967} Ostlund further classified the spin singlet and triplet instabilities into real and imaginary ones, and discussed relation between RHF using complex-valued orbitals (cRHF) and the spin-singlet instabilities.\cite{ostlund} Seeger and Pople revisited the constraints and stability in the HF theory, and formulated an orbital theory named "general Hartree--Fock (GHF)" for further energy minimization following detection of an instability.\cite{seeger_pople1977} Hirao and Nakatsuji took an approach based on a cluster expansion of a wavefunction towards the HF instabilities, adapting the spin symmetry to the single electronic excitation operator.\cite{hirao-closed,hirao-open} The cluster-expansion approach enabled comprehensive understanding of the HF stability and showed a new concept of "stability dilemma": Even if the variational space is extended, an orbital theory with energy lower than that of RHF cannot be obtained when the RHF wavefunction is stable. Hirao and Nakatsuji proved that both the stability and symmetry dilemmas are solved by an extension of the variational space 
in conjunction with
an appropriate symmetry projection, classifying the existing orbital theories and proposing new ones. Fukutome rigorously examined UHF applied to many practical cases in chemistry and enabled intuitive understanding of UHF in the context of chemical properties.\cite{fukutome} More than 30 years later, Yamada and Hirata introduced a new concept named "form degeneracy" and proved existence of the HF instability based on it, enabling detection of such instability and guess (although not prediction) of the instability in each chemical system.\cite{yamada-hirata}. 
As examples, Yamada and Hirata explained Jahn-Teller distortion in addition to covalent bond breaking with the spin-singlet and triplet instabilities.\cite{yamada-hirata}
Considering these rigorous works on the HF stability, the capability of the single-reference electronic structure theories proposed to describe strong correlation is expected to be explained if they have correspondence to some orbital theories.

\stano{The explicit treatment of the electronic Coulomb hole, which accounts for short-range electron correlation effects not captured within standard orbital-theory frameworks, also dates back to the 1950s.\cite{lowdin1959}} Kato proved that an asymptotic behavior of the exact wavefunction of a multielectronic system is
\begin{equation}
    \lim_{r_{ij} \to 0}\Big(\frac{\partial\Psi}{\partial r_{ij}}\Big) = \frac{1}{2}\Psi(\mathbf{r}_1 \cdots \mathbf{r}_N), \label{eq:cusp}
\end{equation}
and each electronic structure theory to describe weak correlation should be constructed to satisfy this cusp condition.\cite{kato1957} 
However, traditional configuration-interaction
(CI)
 type wavefunctions, including those of practical CC theories such as CCSD\cite{ccsd} and that with perturbative triples called CCSD(T)\cite{ccsdtpert1,ccsdtpert2}, converge too slowly to the wavefunctions satisfying the cusp condition 
(\ref{eq:cusp})
with respect to the basis set. The most widely accepted solution to this convergence limit matter is including explicit $r_{12}$ dependence in the wavefunction as Hylleraas pointed 
out
in 1929.
\cite{kutzelnigg1985,noga1992,noga-kutzelnigg,noga-klopper-kutzelnigg1997,Klopper-Noga2003} 
\jozef{In R12/F12 theories stemming from the original Ansatz of Kutzelnigg\cite{kutzelnigg1985} , the traditional configuration state function space is extended
by functions created by the action of a correlation factor (operator)
$\hat{\mathcal{F}}$  on a reference Slater determinant $|\Phi_0\rangle$, from which
the content described by the conventional CI expansion is outprojected. As a result
\bea
    |\Psi\rangle &=& {\hat{\mathcal{R}}}^{\prime} |\Phi_0\rangle +
|\Psi_{\mathrm{CI-type}}\rangle   \label{eq:F12andCI} \\
{\hat{\mathcal{R}}}^{\prime} &=& (1-\hat{\mathcal{M}})\hat{\mathcal{F}} \,.
\eea
$\hat{\mathcal{F}} \equiv \sum_{p>q}\hat{\mathcal{F}}(r_{pq})$
is a function of the inter-electronic distances $r_{pq}$ taking care of
the correct description of the cusp condition.\cite{kato1957}
Treating it in the normal ordered second quantization formalism, action of ${\hat{\mathcal{R}}}^{\prime}$ on $|\Phi_0\rangle$ formally creates 
single and double excitations to the complete virtual space (i.e. virtual space corresponding to the "complete" basis) from which the excitations 
covered by the second part of Eq. (\ref{eq:F12andCI}) are skipped. Formally, those excited configurations are then "externally" contracted, 
which means that for each pair of occupied orbitals from $|\Phi_0\rangle$ the double excitations are summed up, and similarly the single excitation 
for each occupied orbital.  This gives rise to the creation of "pseudo-excitations" when a pair of occupied orbitals in $|\Phi_0\rangle$ is formally replaced 
by the product of the correlation factor with the pertinent (same) orbital pair. However, using these "pseudo-excitations" in subsequent F12 methods 
would give rise to an approach that would not be invariant with respect to rotations among occupied orbitals. An orbital invariant treatment was suggested by Klopper,\cite{KLOPPER1991-orbinv} 
in which the pair of occupied orbitals in $|\Phi_0\rangle$ may be formally replaced by the product of the correlation factor with any other orbital pair.
If one defines a pseudo excitation operator $\hat{\mathcal{R}}$, stemming from ${\hat{\mathcal{R}}^{\prime}}$ as aforementioned,
$\hat{\mathcal{R}}$ can be introduced in the exponential expansion within the coupled cluster ansatz\cite{noga1992,noga-kutzelnigg,noga-klopper-kutzelnigg1997} 
There have been many developments of 
CC-R12/F12 towards } 
more efficiency for example those by Ten-no and co-workers\cite{bokhan2008,noga_ten-no2008,bokhan2009,ten-no_noga2012,ten-no2012}, \jozef{Werner and co-workers,\cite{AdlerWHJ2007,KniziaWHJ2009,AdlerWHJ2009}}
Tew and co-workers,\cite{TEW200832,HOFENER200925,Tew2010,kohn_tew201001,kohn_tew201011,haettig-review2012}
as well as those by Nagy, K\'{a}llay and co-workers.\cite{kallay2021,kallay2023,mester-kallay2023,mester-kallay2025,ladoczki} It should be noted that the R12/F12 scheme is also adopted to other electronic structure theories.\cite{kong2012,haettig-review2012} 
The R12/F12 scheme not only accelerates the convergence towards the basis set limit but also recovers major part of weak correlation lost from the conventional CI-type expansion as shown by H\"{o}fener 
\textit{et al.}
\cite{HOFENER200925} Encouraged by this fact, some researchers proposed
to use the F12 theory for describing more accurate weak correlation by single-reference electronic structure theories with strong correlation: Kats applied F12 to the distinguishable-cluster singles and doubles (DCSD).\cite{kats2015} 
Two
of us applied F12 to pCCD and proposed a theory named "pCCSD-F12".\cite{kedzuch2020}

In this work, we first revisit the cluster-expansion approach for closed-shell systems formulated by Hirao and Nakatsuji\cite{hirao-closed}. We then show that one modification of CCSD for closed-shell systems, singlet-paired CCSD named CCSD0\cite{bulik}, corresponds to an orbital theory obtained by the cluster-expansion approach 
\stanoo{and shares its leading paired configurations with the complex molecular orbital framework.}
\stano{This correspondence indicates that while CCSD0 captures the essential features of strong correlation in most closed-shell systems, it inherently lacks a comprehensive description of the remaining correlation energy due to the omission of the same-spin channels. To recover these missing same-spin correlation effects,
} \jozef{we adopt the explicitly correlated F12 framework to CCSD0. We also formulate its perturbative triples correction. 
We test the new theory for nitrogen, carbon dioxide, methylene and cyclobutadiene to check whether
it properly describes both strong and weak correlations in the single-reference scheme.} 

\section{Theory}

In the present work, indices $i,j,\dots,o$ are used for occupied spin orbitals, $a,b,\dots,f$ are for virtual spin orbitals included in the computational basis, and $p,q,r,s$ are for arbitrary spin orbitals in the computational basis. 
Greek letters $\kappa$, $\lambda$, $\mu$, $\nu$ 
represent any spin orbital within the complete basis set and $\alpha$, $\beta$, $\gamma$, $\delta$ 
denote virtual spin orbitals which are complementary to the computational basis.
In explicit formulation of matrix elements 
indices $p'$, $q'$, $\dots$ 
are used
for the auxiliary basis which approximates the complete basis set within 
the
resolution of identity \cite{valeev-cabs} 
and additional spin orbitals in the auxiliary basis set as complement to the computational basis 
are
denoted as $p"$, $q"$, $\dots$, i. e. $i\cup a\cup p" \equiv p\cup  p" \equiv p'$. 
The Einstein summation convention is used throughout unless otherwise specified.
The reference Slater determinant $|\Phi_0\rangle$ representing the Fermi vacuum takes the following standard form:
\begin{eqnarray}
    |\Phi_0\rangle &&= \prod_{i=1}^{n}a_{i\alpha}^{\dagger}a_{i\beta}^{\dagger}|\rangle \nonumber   \\
    &&= \det[\phi_1(\mathbf{r}_1)\alpha(\xi_1)\phi_1(\bar{\mathbf{r}}_1)\beta(\bar{\xi}_1)\cdots\phi_n(\mathbf{r}_n)\alpha(\xi_n)\phi_n(\bar{\mathbf{r}}_n)\beta(\bar{\xi}_n)]
\end{eqnarray}

\subsection{CCSD0-F12Pi theory}
\lbl{subsec:ccsd0-f12pi}

It was already known that spin-adapted CC theories are effective for describing strong correlation by 
early developers
of the CC theory\cite{paldus1977,paldus_adams_cizek1977} before the 
\stanoo{formulation}
by Bulik \textit{et al.}
\cite{bulik} In the CCSD0 
\jozef{framework, the full double-excitation operator space of conventional CCSD is explicitly decoupled into independent spin-singlet and spin-triplet excitation channels, where the volatile same-spin triplet components are systematically omitted to maintain numerical regularization:}
\be
    |\Psi_{\mathrm{CCSD0}}\rangle = \exp{(\hat{T}_1 + \hat{T}^{[0]}_2)}|\Phi_0\rangle, \lbl{eq:CCSD0_ansatz} 
\ee
\stanoo{where $\hat{T}_{1}$ denotes the standard single-excitation operator, and the double-excitation pool is restricted to the singlet-paired space via:}
\be
	\hat{T}^{[0]}_2 = \frac{1}{2} \sigma^{ab}_{ij}\hat{P}^{\dagger}_{ab}\hat{P}_{ij}. \lbl{eq:T2Sig}       
\ee
\stanoo{Here, $\sigma_{ij}^{ab}$ represent the singlet cluster amplitudes, while the spin-adapted creation and annihilation pair operators are explicitly defined as:}
\bea
    \hat{P}_{ij} &=& \frac{1}{\sqrt{2}} (a_{j\alpha}a_{i\beta}+a_{i\alpha}a_{j\beta}), \\ \lbl{eq:Pij_def} 
    \hat{P}^{\dagger}_{ab} &=& \frac{1}{\sqrt{2}} (a^{\dagger}_{a\beta}a^{\dagger}_{b\alpha}+a^{\dagger}_{b\beta}a^{\dagger}_{a\alpha}). \lbl{eq:Pab_def}
\eea
\stanoo{By truncating the volatile triplet excitation pathways, the parent CCSD0 equations secure global qualitative stability across the potential energy surface. However, this truncation inherently sacrifices the physically real same-spin dynamic electron correlation.}

The explicitly correlated R12/F12 function is given by Eq. (\ref{eq:F12andCI}) utilizing the $\hat {\mathcal F}$ operator. 
The normal-ordered form of the $\hat {\mathcal F}$ operator with respect to the reference Slater determinant (Fermi vacuum) $\ket {\Phi_0} $ is given by
\be
\hat {\mathcal F}_N= \sum_{p>q} {\mathcal F}(r_{pq}) - \langle \Phi_0 \vert \sum_{p>q} {\mathcal F}(r_{pq}) \vert \Phi_0
\rangle
= \quat \bar F_{\mu \nu}^{\kappa \lambda}
\tilde a_{\kappa\lambda}^{\mu\nu} 
+ {\bar F}_{\mu i}^{\kappa i} \tilde a_{\kappa}^{\mu} .
\label{eq:rN}\ee
Similarly, the normal-ordered form of the exact non-relativistic Hamiltonian is given by
\bea
\hat H_N &=& \hat H - \bra  {\Phi_0} \hat H \ket {\Phi_0} =
\hat F_N + \hat W_N \,,   \label{eq:HN} \\
\hat F_N &=&  f_{\kappa}^{\lambda}
\tilde a^{\kappa}_{\lambda} = \left[(\half\nabla^2)  _{\kappa}^{\lambda}
 + v_{\kappa}^{\lambda}\right]\tilde a^{\kappa}_{\lambda} +
\bar g_{{\kappa} i }^{{\lambda} i } \tilde a^{\kappa}_{\lambda}\,,\label{eq:FN}\\
\hat W_N &=& \textstyle{1\over 4}\bar g_{\kappa\lambda}^{\mu\nu}
{\tilde a}^{\kappa\lambda}_{\mu\nu}\,, \label{eq:WN}
\eea
where $v$ is the one particle potential and $\hat g$ is the two electron repulsion operator. 
The Fock operator is split into the Hartree ($h$) and exchange ($k$) parts like $\hat f = \hat h - \hat k$, where
\bea \hat h = h_{\kappa}^{\lambda}\tilde a^{\kappa}_{\lambda}= \left[
(\half\nabla^2)_{\kappa}^{\lambda}+v_{\kappa}^{\lambda} +g_{{\kappa} i }^{{\lambda} i }\right] \tilde a^{\kappa}_{\lambda}\,;&\ \  &
\hat k = k_{\kappa}^{\lambda}\tilde a^{\kappa}_{\lambda} = g_{{\kappa} i}^{i{\lambda}} \tilde a^{\kappa}_{\lambda}\,.\eea

\jozef{Action of $\hat {\mathcal F}_N$ on $\vert \Phi _0 \rangle$ yields\cite{noga-klopper-kutzelnigg1997}}
\be \hat \mF _N \vert \Phi_0 \rangle =
\left(
\quat \bar F^{ij}_{ab}\tilde a_{ij}^{ab} +\quat \bar F^{ij}_{\alpha\beta}\tilde a_{ij}^{\alpha\beta}
+ \half \bar F^{ij}_{a\beta}\tilde a_{ij}^{a\beta}
+ \bar F^{ij}_{aj}\tilde a_i^a +\bar F^{ij}_{\alpha j}\tilde a_i^{\alpha}
\right)\vert \Phi_0 \rangle \,. \label{eq:rNPhi}\ee
\stanoo{To prevent the double-counting of configuration pathways between the conventional wave function layer and the explicitly correlated expansion, the conventional cluster excitations must be outprojected from the initial framework. Crucially, within the regularized single-reference scheme, this outprojection operator $\hat{\cal{M}}$ isolates exclusively the spin-singlet degrees of freedom that remain active in the underlying coupled-cluster equations} 
\be \big(1-\hat {\cal M}
) \hat F_N \ket{\Phi_0}
= \left(
\quat \bar F^{ij}_{\alpha\beta}\tilde a_{ij}^{\alpha\beta}
+ \half \bar F^{ij}_{\alpha b}\tilde a_{ij}^{\alpha b}
+ \quat \bar F^{ij}_{ab}\tilde a_{ij}^{ab}(Pi)
+\bar F^{ij}_{\alpha j}\tilde a_i^{\alpha}
\right)\vert \Phi_0 \rangle \,, \label{eq:rNPhi-proj}
\ee
\stanoo{where the inclusion of the $(Pi)$ suffix directly targets the supplementary restoration of the same-spin triplet channels;
\jozef{\footnote{The specific labels (Sig) and (Pi) are adopted here directly from the original cluster regularization terminology of Bulik et al.\cite{bulik} to maintain baseline compatibility, rather than denoting spatial molecular orbital symmetries.}}
this explicit complement is necessary because the conventional projection layer inside $\hat{\cal{M}}$ within the CCSD0 framework 
is rigorously restricted to the singlet configuration space according to}
\stano{
\be
\hat {\cal M} = \ket {\Phi_0} \bra {\Phi_0} +\tilde a_i^a \vert \Phi_0 \rangle \langle  \Phi_0\vert  \tilde
a^i_a +\tilde a_{ij}^{ab}(Sig) \vert \Phi_0 \rangle \langle  \Phi_0\vert  \tilde
a^{ij}_{ab}(Sig) + \dots \,.
\ee
}
\stanoo{Here, the label $(Sig)$ designative of the double-excitation strings specifies that only the singlet-paired, opposite-spin amplitudes enter the active iterative optimization loop.}
The following operator is defined
\bea
\hat \cR &=& \hat \cR _1 + \hat \cR _2 = c_k^i\tilde{\cR}_{i}^{k} + \quat c^{ij}_{kl}
\tilde{\cR}_{ij}^{kl} \lbl{eq:cR}\,, \\
\tilde{\cR}_{i}^{k} &=&
 \bar F_{\alpha j}^{kj}\tilde a_{i}^{\alpha}\,,
 \lbl{eq:pseudo1}\\
\tilde{\cR}_{ij}^{kl} &=&
\half\bar F_{\alpha\beta}^{kl}\tilde a_{ij}^{\alpha\beta} +
\bar F_{\alpha b}^{kl}
\tilde a_{ij}^{\alpha b} 
+ \half\bar F_{ab}^{kl}\tilde a_{ij}^{ab}(Pi) \,, 
 \lbl{eq:pseudo2Pi}
 \eea
where $\tilde{\cR}_{i}^{k}$ and $\tilde{\cR}_{ij}^{kl}$ are 
pseudo single and double-excitation operators, respectively,
while
$c_k^i$ and $c^{ij}_{kl}$ are F12 amplitudes. 
\stanoo{The suffix $(Pi)$ assigned to the final term in Eq. (\ref{eq:pseudo2Pi}) underscores the direct restoration of the spin-triplet pseudo-excitation pathways. 
Rather than acting as a conventional completeness expansion tool, this specific same-spin component functions as a targeted correlation supplement. 
By capturing the same-spin contributions directly via the interelectronic coordinate $r_{12}$, the F12 layer seamlessly fills the correlation deficit 
caused by the parent regularization without expanding the parameter space or reintroducing unconstrained cluster degrees of freedom. 
This unique physical interplay between the singlet-restricted conventional series and the triplet-complementary explicit layer defines 
the core architecture of the present methodology, which we designate as "CCSD0-F12Pi".
Note that the $(Pi)$ pseudo-excitations couple triplet-paired electron and hole pairs to an overall singlet;
the closed-shell reference therefore retains its singlet character by construction.}

\stanoo{It is important to emphasize that the standard fixed geminal amplitudes remain appropriate for the triplet channels, even though the $\hat {R}$ operator explicitly spans pseudoexcitations into the computational virtual space. 
The physical validity of applying these cusp-determined fixed values within a truncated virtual expansion layer is firmly established by the underlying orthogonalization framework of Noga and Kutzelnigg\cite{noga-kutzelnigg}, 
and has been successfully verified in our previous pCCSD-F12 molecular applications\cite{kedzuch2020}.}
\stanoo{Furthermore, this framework is fundamentally consistent with standard explicitly correlated treatments; in the routine implementation of the conventional \jozef{SP ansatz\cite{Ten_no2004}}, 
the standard pseudo-double excitation operator inherently utilizes a companion layer—specifically matching the second term in Eq. (\ref{eq:pseudo2Pi})—which explicitly spans 
double-excitation pathways into the computational virtual orbital space. Since the fixed universal prefactors seamlessly manage these internal computational configurations 
in standard applications, their extension to the structurally analogous same-spin $(Pi)$ channel in Eq. (\ref{eq:pseudo2Pi}) remains conceptually consistent and methodologically sound.}

Now the wavefunction is written like that of the standard CCSD-F12 one as
\bea 
\ket {\Psi} & = &  e^{\hat S} \ket {\Phi_0} \,, \lbl{eq:CCansatz} \\
{\hat S} & = & {\hat T}+\hat {\mathcal R}\,. \lbl{eq:S}
\eea

Similarity-transformed effective Hamiltonians are given with (\ref{eq:pseudo2Pi}) as follows:
\bea 
\bar H _\cR & = & \exp(-\hat \cR )\hat H_N   \exp(\hat \cR )  \lbl{eq:HRsim}\\
\bar H _T & = & \exp(-\hat T )\hat H_N   \exp(\hat T )\,.  \lbl{eq:HTsim} 
\eea
For evaluation of the energy and excitation amplitudes, the following form of the Schr\"odinger equation is used:
\be \exp(-\hat S )\hat H _N  \exp(\hat S) \ket{\Phi_0} = E \ket{\Phi_0} \,, \lbl{eq:SE} \ee
In the following subsections, explicit forms of the equations are presented which provide the energy and excitation amplitudes in the same manner as the standard CCSD-F12 method \cite{noga_ten-no2008}.

\subsubsection{Energy}
\label{subsub:energy}

\jozef{By projecting Eq. (\ref{eq:SE}) onto the reference, the energy is obtained as}
\be
E = \bra {\Phi_0} \bar H _T \ket {\Phi_0} + \quat \cV_{ij}^{kl} c^{ij}_{kl}
-\cV_{i}^{k} c^{i}_{k} + \cG_{ij}^{kl} c^{i}_{k}c^{j}_{l} -
\cU_{ij}^{ka}c^{i}_{k}t^j_a \,, \lbl{eq:Enexp}
\ee
where the first term is the conventional CC energy, and
\be
 \cV_{pq}^{kl} = (\overline{F_{12}/r_{12}} )_{pq}^{kl} - \bar g_{pq}^{mq"}\bar F_{mq"}^{kl} 
 - \half \bar g_{pq}^{rs}\bar F_{rs}^{kl} + \half \bar g_{pq}^{ab}\bar F_{ab}^{kl}(Pi) \,. \lbl{eq:Vaux}
\ee
For singlet pairs of pseudo-excitations, the last term of Eq. (\ref{eq:Vaux}) is absent and the form of $\cV_{pq}^{kl}$ is the same as that for the standard R12/F12 theory which is described in detail elsewhere.\cite{SKappC05}
For triplet pairs, the final form of $\cV_{pq}^{kl}$ is as follows:
\bea
 \cV_{pq}^{kl}& =& (\overline{F_{12}/r_{12}} )_{pq}^{kl} - \bar g_{pq}^{mq"}\bar F_{mq"}^{kl}
 - \half \bar g_{pq}^{nm}\bar F_{nm}^{kl} - \half \bar g_{pq}^{nb}\bar F_{nb}^{kl} - \half \bar g_{pq}^{am}\bar F_{am}^{kl} - \half \bar g_{pq}^{ab}\bar F_{ab}^{kl}
 + \half \bar g_{pq}^{ab}\bar F_{ab}^{kl} \nonumber \\
 &=& (\overline{F_{12}/r_{12}} )_{pq}^{kl} - \bar g_{pq}^{mq"}\bar F_{mq"}^{kl}
 - \half \bar g_{pq}^{nm}\bar F_{nm}^{kl} - \bar g_{pq}^{nb}\bar F_{nb}^{kl}. \lbl{eq:Vexpl}
\eea
The last three terms of Eq. (\ref{eq:Enexp}) are common to the standard R12/F12 theory and can be evaluated within auxiliary basis set, which is orthogonal to the computational basis set, introduced by Valeev\cite{valeev-cabs} as \shota{complementary auxiliary basis set} (CABS) approximation:
\bea
 \cV_{p}^{i} &=&  f _{p}^{\alpha} F ^{i}_{\alpha } \eqaux f _{p}^{p"} F ^{i}_{p"} \,,
 \lbl{eq:V1}\\
 \cG_{pq}^{ij} &=&  \bar g _{pq}^{\alpha \beta}  F ^{i}_{\alpha } F^{j}_{\beta} \eqaux \bar g _{pq}^{p"q"}  F ^{i}_{p"} F^{j}_{q"}\,,
 \lbl{eq:G2}\\
 \cU_{pq}^{ir} &=&  \bar g _{pq}^{\alpha r}  F ^{i}_{\alpha } \eqaux \bar g _{pq}^{p" r}  F ^{i}_{p"}\,.
 \lbl{eq:U2}
\eea

\subsubsection{$\hat T$ amplitudes}

The $t$-amplitude equations are obtained by projecting excited-state configurations onto Eq. (\ref{eq:SE}) as follows:
\bea
0 &=& (\bar H_T )^i_a 
+(\bar H_\cR ')^{i}_{a} - (\bar H_\cR ')^{i}_{k} t^k_a  
+(\bar H_\cR ')^{c}_{a}t_c^i + (\bar H_\cR ')^{c}_{k} (t_{ac}^{ik} - t_a^k t_c^i)
\nonumber \\ 
&& +(\bar H_\cR ')^{ic}_{ak} t_c^k - (\bar H_\cR ')^{ic}_{lk} (t_a^l t_c^k + \half t_{ac}^{lk})  
\,, \lbl{eq:T1exp}\\
0 &=& (\bar H_T )^{ij}_{ab}
+ (\bar H_\cR ')^{ij}_{ab} - (\bar H_\cR ')^{ij}_{kb} t_a^k - (\bar H_\cR ')^{ij}_{al} t_b^l + (\bar H_\cR ')^{ij}_{kl} (t_a^k  t_b^l + \half t_{ab}^{kl})
\nonumber \\ 
&& - (\bar H_\cR ')^{k}_{m} t_{ab}^{ml}\bar \delta_{kl}^{ij}
+ [(\bar H_\cR ')^{e}_{c} t_{ed}^{ij}      
-(\bar H_\cR ')^{ije}_{ckl}(t_{de}^{kl}+t^k_dt^l_e) ]\bar\delta_{ab}^{cd}
+(\bar H_\cR ')^{ijc}_{abl}t^l_c
\,, \lbl{eq:T2exp}
\eea
where $\bar H_{\cR}'$ is a shifted effective Hamiltonian; $ \bar H_{\cR}' = \bar H_{\cR} - \hat H_N $.

\bea
(\bar H_\cR ')^i_{p\phantom {qr}} &=& \half \cV_{pn}^{kl} c_{kl}^{in}  - \cV_p^k c_k^i
- \cU_{mp}^{ki} c^m_k + \cG_{mp}^{kl} c_k^m c_l^i  \nonumber \\
&& -\cV^{kli}_{amn}\delta^a_p c_{kl}^{nm}
+ \cY^{klm}_{ano}\delta^a_p c_{kl}^{in}c^o_m \,, \lbl{eq:HcRip1}  \\
(\bar H_\cR ')^c_{p\phantom {qr}} &=& 
- \cU_{mp}^{kc} c^m_k -
\cV^{klc}_{amn}\delta^a_p c_{kl}^{nm} \,, \lbl{eq:HcRap1} \\
(\bar H_\cR ')^{ic}_{pq\phantom {r}} &=&
-\cU_{pq}^{kc} c^i_k + \half\cV^{klc}_{aor} c_{kl}^{io}\bar \delta_{pq}^{ar}\,,  \lbl{eq:HRicpq} \\
(\bar H_\cR ')^{ij}_{pq\phantom {r}} &=& \half \cV_{pq}^{kl} c_{kl}^{ij}
+( \cU_{pq}^{km} c^n_k 
+ \cG_{pq}^{kl} c_k^m c_l^n ) \bar \delta_{mn}^{ij} \nonumber \\
&&
+ \half (\cV^{kln}_{aor}-\cY^{kll_1}_{aor}c_{l_1}^n) c_{kl}^{mo}\bar \delta_{pq}^{ar}  \bar \delta_{mn}^{ij} \nonumber  \\
&&
+\quat \cV^{kk_1l_1l}_{ak_2l_2b}c_{kk_1}^{mk_2}c_{l_1l}^{l_2n}\bar \delta_{pq}^{ab} 
 \bar \delta_{mn}^{ij} 
-\half \cY^{klm}_{arn}c_{m}^n c_{kl}^{ij}\bar \delta_{pq}^{ar}\,, \lbl{eq:HRijpq} \\
(\bar H_\cR ')^{ijc}_{pqr} &=& \quat  \cV^{mnc}_{atu} 
 c_{mn}^{ij} \bar \delta ^{atu}_{pqr}\,. \lbl{eq:HcRijcpqr} 
\eea

\bea
\cV_{apq}^{ijr} &=&  \bar g _{pq}^{\beta  r} \bar F ^{ij}_{a \beta} + \bar g _{pq}^{b  r} \bar F ^{ij}_{a b} (Pi) \,,
  \lbl{eq:V3}\\
\cY_{apq}^{ijk} &=&  \bar g _{pq}^{\beta  \alpha} \bar F ^{ij}_{a \beta} F ^{k}_{\alpha } 
+ \bar g _{pq}^{b  \alpha} \bar F ^{ij}_{a b} (Pi)
F ^{k}_{\alpha }\,,  
 \lbl{eq:Y3}\\
\cV_{apqb}^{ijkl} &=&  
  \half \bar g _{pq}^{\beta  \alpha} \bar F ^{ij}_{a \beta} \bar F ^{kl}_{\alpha b} 
+ \half \bar g _{pq}^{d  \alpha} \bar F ^{ij}_{a d} (Pi) \bar F ^{kl}_{\alpha b}
+ \half \bar g _{pq}^{\beta  c} \bar F ^{ij}_{a \beta} \bar F ^{kl}_{c b} (Pi)
+ \half \bar g _{pq}^{d  c} \bar F ^{ij}_{a d} (Pi) \bar F ^{kl}_{c b} (Pi)
\,.
  \lbl{eq:V4}
\eea

\subsubsection{$ \hat \cR $ amplitudes}

\stano{While most \shota{of the} contemporary F12 implementations utilize the computationally fixed SP-ansatz\cite{bokhan2008} framework, we present the explicit residual equations for the F12 pseudo-excitation amplitudes here for the sake of completeness of the theory and to detail our numerical framework. Crucially, both the SP-ansatz scheme and the iterative optimization of the amplitudes have been implemented in our pilot software and evaluated in the benchmark section. 

The working equations for the determination of the $ \hat \cR $ amplitudes are derived by projecting the similarity-transformed Schrödinger equation  (\ref{eq:SE}) onto the pseudo-excited configurations, namely}
$\bra {\Phi_0} (\cR^{\dagger})_{i}^{k}$ and $\bra {\Phi_0} (\cR^{\dagger})_{ij}^{kl}$.  

F12 single pseudo-excitation amplitude equations read:
\bea
0&=&-\cB_k^l c_l^i  + (\cV ^{\dagger}) ^i_k + (\cV ^{\dagger}) ^a_k t_a^i + [(\cU ^{\dagger}) ^{ic}_{km} + (\cU ^{\dagger}) ^{ac}_{km} t_a^i]t^m_c  +(\cZ _{km}^{il} + \cZ _{km}^{al} t_a^i) c_l^m  \nonumber \\  
&&+ \quat (\cE ^{loi}_{kmn}  + \cE ^{loa}_{kmn} t_a^i)c_{lo}^{mn}
- \cZ _{km}^{cl} c_l^i t_c^m + \cX ^l_k c_l^n (\bar H_N)_n^i +\half (\cD _{kl}^{mn} + \cQ_{kl}^{mn} + \cL ^{mn}_{kl}) c_{mn}^{il}   \nonumber \\  
&&+ \half (\cE^{jnc}_{klm}t_{c}^{m} - \cD ^{jno}_{klm} c_o^m) c_{jn}^{il}
\,, \lbl{eq:C1}
\eea
where 
\be
(\bar H_N)^i_k = (\bar H_T)^i_k +(\bar H'_\cR)^i_k -\cU_{kl}^{ma}c_m^i t_a^l -\cU_{lk}^{ma}c_m^l t_a^i \,,
\ee
and
matrices $\cV ^k_i $ and $ \cU ^{km}_{ic} $ 
are
defined in Section \ref{subsub:energy}.  
The matrices
\bea
\cB_{k}^{i} &=&   F _{k}^{\alpha}  f^{\beta}_{\alpha} \bar F ^{i}_{\beta} \\
\cZ^{ip}_{kq} &=& F_k^{\alpha} \bar g_{\alpha q}^{\beta p} F ^{i}_{\beta} \\
\cE_{kmn}^{ijp}&=& F_k^{\alpha} (\bar g^{\beta p}_{mn} \bar F^{ij}_{\alpha\beta} + \bar g^{bp}_{mn} \bar F^{ij}_{\alpha b} ) \\
\cX_{k}^{i} &=&  F _{k}^{\alpha } F ^{i}_{\alpha} \\
\cD_{kl}^{ij} &=& F_k^{\alpha} \bar g_{\alpha l}^{\beta\gamma}   F ^{i}_{\beta}F^j_{\gamma} \\
\cQ_{kl}^{ij} &=& F_k^{\alpha} (f_l^{\gamma}\bar F_{\alpha\gamma}^{kl}+f_l^{c}\bar F_{\alpha c}^{kl})  \\
\cD_{lmn}^{ijk} &=& F_l^{\alpha} (\bar g^{\beta\gamma}_{mn} \bar F^{ij}_{\alpha\beta} + \bar g^{b\gamma}_{mn} \bar F^{ij}_{\alpha b} ) F_{\gamma}^k 
\eea
are 
common to the
standard CC-F12 method. 

In addition to CCSD-F12\cite{noga_ten-no2008}, the matrix
$ \cL ^{mn}_{kl} $ 
involves a
term due to
spin-triplet pair
pseudo-double excitations
\be
\cL ^{mn}_{kl} = F_k^{\alpha} (\half \bar g_{\alpha l}^{\beta\gamma} \bar F_{\beta\gamma}^{kl} + \bar g_{\alpha l}^{\beta c} \bar F_{\beta c}^{kl} + \half \bar g_{\alpha l}^{bc} \bar F_{bc}^{kl}(Pi) ) = F_k^{p"}\cV _{p" l}^{kl} \,,
\ee
where
$\cV_{pq}^{kl} $ 
are already derived in Eqs. (\ref{eq:Vaux}) and (\ref{eq:Vexpl}).

Pseudo-double excitation amplitudes are determined by solving the following set of equations:
\bea
0 &=&\half\cB _{kl}^{mn} c_{mn}^{ij} +(\cV^{\dagger}) _{kl}^{ij} + (\cV^{\dagger}) _{kl}^{aj}t_a^i + (\cV^{\dagger}) _{kl}^{ib}t_b^j + (\cV^{\dagger}) _{kl}^{ab}(t_a^i t_b^j +\half t_{ab}^{ij}) 
+ \quat \cX _{kl}^{o_1 o_2} c_{o_1 o_2}^{mn}(\bar H_N)_{mn}^{ij} + \half \cP _{kl}^{mn} c_{mn}^{ij} 
\nonumber \\
&&- \half [\cX _{kl}^{o_1 o_2} c_{o_1 o_2}^{mo} (\bar H_N)_{o}^{n}+ (\cZ _{klo}^{o_1 o_2 n} + \cZ _{klo}^{o_1 o_2 a} t_a^n) c_{o_1 o_2}^{om}] \bar \delta_{mn}^{ij} +\half \cZ _{klo}^{mna} c_{mn}^{ij} t_a^o
+ (\cY^{\dagger})_{kl}^{ab} t_{ab}^{ij}
\nonumber \\
\noalign{\smallskip}
&& -[ (\cV^{\dagger})_{klo}^{anc} + (\cV^{\dagger})_{klo}^{abc} t_b^n ] t_{ac}^{om} \bar \delta _{mn}^{ij}  
+ [(\cV^{\dagger})_{klo}^{aij} + (\cV^{\dagger})_{klo}^{abj} t_b^i + (\cV^{\dagger})_{klo}^{aic} t_c^j + (\cV^{\dagger})_{klo}^{abc} (t_b^i t_c^j + \half t_{bc}^{ij}) ] t_a^o 
\nonumber \\
\noalign{\smallskip}
&&- [(\cE^{\dagger})_{klm}^{nij} + (\cE^{\dagger})_{klm}^{nbj} t_b^i + (\cE^{\dagger})_{klm}^{nic} t_c^j + (\cE^{\dagger})_{klm}^{nbc} (t_b^i t_c^j + \half t_{bc}^{ij}) ] c_n^m
- \cK _{klo}^{anb} c_n^o t_{ab}^{ij} + \cM _{klo}^{amn} c_{mn}^{ij} t_a^o 
\nonumber \\
\noalign{\smallskip}
&& + [\cK _{klo}^{ao_1b} c_{o_1}^n t_{ab}^{mo} - (\cK _{klo}^{ao_1m} c_{o_1}^n + \cK _{klo}^{ao_1b} c_{o_1}^n t_b^m) t_a^o + (\cJ_{klo}^{o_1 o_2 m} + \cJ_{klo}^{o_1 o_2 c} t_c^m ) c_{o_1 o_2}^{no}]\bar \delta _{mn}^{ij}
\nonumber \\
\noalign{\smallskip}
&&+ (\cI _{okl}^{mno_1} c_{o1}^o + \cI _{kl}^{mn}) c_m^i c_n^j + [(\cL^{\dagger})_{kl}^{on} c_o^m + (\cL^{\dagger})_{kl}^{ob} c_o^m t_b^n - \half \cL _{klo}^{o_1o_2o_3} c_{o_1o_2}^{mo} c_{o_3}^n] \bar \delta _{mn}^{ij} + \half \cL _{klo}^{mno_1} c_{mn}^{ij} c_{o_1}^o   
\nonumber \\
\noalign{\smallskip}
&& - \half (\cP _{olk}^{mno} + \cP _{olk}^{mnc} t_c^o ) c_{mn}^{ij} +\quat\cW _{kl}^{oo_1} c_{oo_1}^{mn} (\bar H_N)_{mn}^{ij}  - \cX _{kl}^{nb} c_n^m (\bar H'_T)_{mb}^{ij}
+ \half \cX _{kla}^{mnb} c_{mn}^{ij} (H'_T)^a_b
\nonumber \\
\noalign{\smallskip}
&& +\eighth   \cZ _{ko_1o_2l}^{o_3o_4o_5o_6} c_{o_3o_4}^{mo_1} c_{o_5o_6}^{o_2n} \bar \delta_{mn}^{ij} +\eighth \cZ _{knml}^{o_1o_2o_3o_4} c_{o_1o_2}^{ij} c_{o_3o_4}^{mn}
- \half \cM_{klom}^{ano_1o_2} c_n^j c_{o_1o_2}^{im} t_a^o  
+ \quat \cJ _{knml}^{o_1o_2o_3o_4} c_{o1}^m c_{o_2}^n c_{o_3o_4}^{ij} 
\nonumber \\
\noalign{\smallskip}
&&+ \quat \cJ _{ko_1o_2l}^{m_1n_1m_2n_2}c^{o_1}_{m_1}c^{m}_{n_1}c_{m_2n_2}^{o_2n} \bar \delta_{mn}^{ij}
+ \half \cK _{knml}^{o_1o_2ab} c_{o_1o_2}^{mn}  t_{ab}^{ij}
+ \quat \cK _{kmnl}^{o_1o_2ab} c_{o_1o_2}^{ij} t_{ab}^{mn}
- \half \cK _{ko_1o_2l}^{o_3o_4ab} c_{o_3o_4}^{mo_1} t_{ab}^{o_2n} \bar \delta_{mn}^{ij}\,,
\lbl{eq:C2} 
\eea
where 
\be
(\bar H_N)^{ij}_{kl} = (\bar H_T)^{ij}_{kl} +(\bar H_\cR)^{ij}_{kl} -\cU_{kl}^{oa}c_o^mt_a^n \bar \delta_{mn}^{ij}\,,
\ee

and
\bea
\cB _{kl}^{ij} & = & \bar F _{kl}^{\alpha \beta} f^{\gamma}_{\alpha} \bar F ^{ij}_{\gamma \beta}
+\bar F _{kl}^{\alpha b} f^{\gamma}_{\alpha} \bar F ^{ij}_{\gamma b}
+\bar F _{kl}^{\alpha b} f^{c}_{b} \bar F ^{ij}_{\alpha c} \nonumber\\  
&&+(\bar F _{kl}^{\alpha \beta} f_{\alpha}^{c} \bar F ^{ij}_{c \beta} + \bar F _{kl}^{\alpha b} f_{\alpha}^{c} \bar F ^{ij}_{cb}(Pi) + H.\,c.) 
+ \bar F _{kl}^{ab}(Pi) f^{c}_{a} \bar F ^{ij}_{cb}(Pi)\,,
\lbl{eq:B2}
\eea
where $ H.\,c. $ means hermitian conjugates.

For derivation of $ \cB _{kl}^{ij} $ 
Approximation C of Ref.\citenum{SKappC05} is used
giving rise to:
\bea B_{ij}^{kl} &\eqaux&
\Lambda_{ij}^{kl} + \frac{1}{2}\left[\bar{(F^2)}_{ij}^{p'l}(f+k)_{p'}^k +
\bar{(F^2)}_{ij}^{kp'}(f+k)_{p'}^l + H.\,c.\right] \nonumber \\
&& - \bar F_{ij}^{p'q'}k_{p'}^{r'}\bar F_{r'q'}^{kl} 
-\bar F_{ij}^{p'm}f_{p'}^{r'} \bar F_{r'm}^{kl}
 +\bar F_{ij}^{mq"}f_m^n \bar F_{nq"}^{kl}
- \bar F_{ij}^{pb}f_p^r \bar F_{rb}^{kl} \nonumber \\
&& - \left(\bar F_{ij}^{mq"}f_m^{p'} \bar F_{p'q"}^{kl}
+ \bar F_{ij}^{p"b}f_{p"}^q \bar F_{qb}^{kl} + H.\,c.\right) \nonumber \\
&& + \bar F _{kl}^{ab}(Pi) f^{c}_{a} \bar F ^{ij}_{cb}(Pi) + (\bar F _{kl}^{p" b} f_{p"}^{c} \bar F ^{ij}_{cb}(Pi) + H.\,c.) \,,
\label{eq:B2aux}
\eea
where $\hat \Lambda(r_{12}) = \gamma^2 exp(-2\gamma r_{12}) $.

After simple manipulations 
following formula is obtained:
\bea B_{ij}^{kl} &\eqaux&
\Lambda_{ij}^{kl} + \frac{1}{2}\left[\bar{(F^2)}_{ij}^{p'l}(f+k)_{p'}^k +
\bar{(F^2)}_{ij}^{kp'}(f+k)_{p'}^l + H.\,c.\right] \nonumber \\
&& - \bar F_{ij}^{p'q'}k_{p'}^{r'}\bar F_{r'q'}^{kl} 
-\bar F_{ij}^{p'm}f_{p'}^{r'} \bar F_{r'm}^{kl}
 +\bar F_{ij}^{mq"}f_m^n \bar F_{nq"}^{kl}
- \bar F_{ij}^{mb}f_m^n \bar F_{nb}^{kl} - (\bar F_{ij}^{mb}f_m^c \bar F_{cb}^{kl} + H.\,c.) \nonumber \\
&& - \left(\bar F_{ij}^{mq"}f_m^{p'} \bar F_{p'q"}^{kl}
+ \bar F_{ij}^{p"b}f_{p"}^n \bar F_{nb}^{kl} + H.\,c.\right)\,.
\label{eq:B2sub}
\eea

With canonical reference using Standard approximation (SA) \cite{WKu91R12gen} 
the final formula is obtained as follows:
\bea B_{ij}^{kl} &\eqsa&
\Lambda_{ij}^{kl} + \frac{1}{2}\left[\bar{(F^2)}_{ij}^{pl}(f+k)_{p}^k +
\bar{(F^2)}_{ij}^{kp}(f+k)_{p}^l + H.\,c.\right] \nonumber \\
&& - \bar F_{ij}^{pq}k_{p}^{r}\bar F_{rq}^{kl} 
-\bar F_{ij}^{an} \bar F_{an}^{kl} \epsilon _a - \bar F_{ij}^{mq} \bar F_{mq}^{kl} \epsilon _m
\label{eq:B2sa}
\eea

Derivation of matrix $ X_{ij}^{kl} $ is 
analogous to the
derivation of $ V_{ij}^{kl} $ matrix 
(\ref{eq:Vaux})
\be
\cX_{ij}^{kl} = (\overline{F_{12}^2} )_{ij}^{kl} - \bar F_{ij}^{mq"}\bar F_{mq"}^{kl}
- \half \bar F_{ij}^{nm}\bar F_{nm}^{kl} - \bar F_{ij}^{nb}\bar F_{nb}^{kl}\,.
\ee

\bea
\cP_{kl}^{ij} &=& 
 \quat \bar F _{kl}^{\alpha \beta}\bar g_{\alpha \beta}^{\gamma \delta} \bar F ^{ij}_{\gamma \delta}
+(\half \bar F _{kl}^{\alpha b}\bar g_{\alpha b}^{\gamma \delta} \bar F ^{ij}_{\gamma \delta} + H.\,c.) 
+      \bar F _{kl}^{\alpha b}\bar g_{\alpha b}^{c \delta} \bar F ^{ij}_{c \delta} \nonumber \\
&& 
+ \quat \bar F _{kl}^{ab}(Pi) \bar g_{ab}^{cd} \bar F ^{ij}_{cd}(Pi) 
+ (\half \bar F _{kl}^{ab}(Pi) \bar g_{ab}^{c\delta} \bar F ^{ij}_{c\delta}
+ \quat \bar F _{kl}^{ab}(Pi) \bar g_{ab}^{\gamma \delta} \bar F ^{ij}_{\gamma \delta} + H.\,c.) \nonumber \\
&\eqaux& 
(\overline{F_{12}^2/r_{12}} )^{ij}_{kl} 
-\half[(\overline{F_{12}/r_{12}} )_{kl}^{pq} \bar F_{pq}^{ij} + H.\,c.]
+\quat \bar F_{ij}^{pq} \bar g_{pq}^{rs} \bar F_{rs}^{kl} - \quat \bar F _{kl}^{ab} \bar g_{ab}^{cd} \bar F ^{ij}_{cd} \nonumber \\ 
&& 
- (\bar F_{kl}^{p"m}  \cV _{p"m}^{ij} - \half \bar F_{kl}^{ab} \cV_{ab}^{ij} + H.\,c.) 
- \bar F_{kl}^{p"m} \bar g_{p"m}^{r"n} \bar F_{r"n}^{ij} \nonumber \\ 
&=& 
(\overline{F_{12}^2/r_{12}} )^{ij}_{kl}
-\half[(\overline{F_{12}/r_{12}} )_{kl}^{pq} \bar F_{pq}^{ij} + H.\,c.] 
\stano{+\quat \bar F_{ij}^{mn} \bar g_{mn}^{oo_1} \bar F_{oo_1}^{kl} + \quat (\bar F_{ij}^{mn} \bar g_{mn}^{cd} \bar F_{cd}^{kl} + H.\,c.)} \nonumber \\
&&
\stano{+\half (\bar F_{ij}^{mn} \bar g_{mn}^{od} \bar F_{od}^{kl} + \bar F_{ij}^{ab} \bar g_{ab}^{od} \bar F_{od}^{kl} + H.\,c.) + \bar F_{ij}^{mb} \bar g_{mb}^{od} \bar F_{od}^{kl} }  \nonumber \\ 
&& 
- (\bar F_{kl}^{p"m}  \cV _{p"m}^{ij} - \half \bar F_{kl}^{ab} \cV_{ab}^{ij} + H.\,c.) 
- \bar F_{kl}^{p"m} \bar g_{p"m}^{r"n} \bar F_{r"n}^{ij} \lbl{eq:Pijkl} \\ 
\cZ^{ijp}_{klm} &=& 
 \bar F_{kl}^{\alpha\gamma} \bar g_{\alpha m}^{\beta p}\bar  F ^{ij}_{\beta\gamma}
+(\bar F_{kl}^{a\gamma} \bar g_{am}^{\beta  p}\bar  F ^{ij}_{\beta\gamma} + H.\,c.)
+\bar F_{kl}^{a\gamma} \bar g_{am}^{b p}\bar  F ^{ij}_{b\gamma}
+\bar F_{kl}^{\alpha c} \bar g_{\alpha m}^{\beta p}\bar  F ^{ij}_{\beta c} \nonumber \\
&&
+ \bar F_{kl}^{ac}(Pi) \bar g_{am}^{b p}\bar  F ^{ij}_{bc}(Pi) 
+(\bar F_{kl}^{ac}(Pi) \bar g_{am}^{\beta p}\bar  F ^{ij}_{\beta c} + H.\,c.) \nonumber \\
&\eqaux&
\half[{(\bar{F}^2)} ^{ij}_{p'l}g^{p'p}_{km} + {(\bar{F}^2)} ^{ij}_{kp'}g^{p'p}_{lm} + H.\,c.] 
-\bar F_{kl}^{p"q"}g_{p"m}^{pr"}\bar F_{r"q"}^{ij}
-\bar F_{kl}^{p"b}g_{p"m}^{pr"}\bar F_{r"b}^{ij} 
-\bar F_{kl}^{aq"}g_{am}^{pb}\bar F_{bq"}^{ij}\nonumber \\
&&
-\big(\bar F_{kl}^{p"q"}g_{p"m}^{pa}\bar F_{aq"}^{ij} + H.\, c.\big)
-\big(\bar F_{kl}^{p"q}g_{p"m}^{pr}\bar F_{rq}^{ij} + H.\, c.\big) 
-\big(\bar F_{kl}^{p"q"}g_{p"m}^{pn}\bar F_{nq"}^{ij} + H.\, c.\big)\nonumber \\
&&
-\bar F_{kl}^{qs}g_{sm}^{rp}\bar F_{qr}^{ij}
+ \bar F_{kl}^{ac} \bar g_{am}^{b p}\bar  F ^{ij}_{bc}
+ (\bar F_{kl}^{ac} \bar g_{am}^{r" p}\bar  F ^{ij}_{r" c} + H.\,c.) \nonumber \\
&=& 
\half[{(\bar{F}^2)} ^{ij}_{p'l}g^{p'p}_{km} + {(\bar{F}^2)} ^{ij}_{kp'}g^{p'p}_{lm} + H.\,c.]
-\bar F_{kl}^{p"q"}g_{p"m}^{pr"}\bar F_{r"q"}^{ij}
-\bar F_{kl}^{p"b}g_{p"m}^{pr"}\bar F_{r"b}^{ij}
-\bar F_{kl}^{aq"}g_{am}^{pb}\bar F_{bq"}^{ij}\nonumber \\
&&
-\big(\bar F_{kl}^{p"q"}g_{p"m}^{pa}\bar F_{aq"}^{ij} + H.\, c.\big)
-\big(\bar F_{kl}^{p"q}g_{p"m}^{pr}\bar F_{rq}^{ij} + H.\, c.\big)
-\big(\bar F_{kl}^{p"q"}g_{p"m}^{pn}\bar F_{nq"}^{ij} + H.\, c.\big)\nonumber \\
&&
\stano{-\bar F_{kl}^{on}g_{nm}^{o_1 p}\bar F_{oo_1}^{ij} - \bar F_{kl}^{cn}g_{nm}^{o_1 p}\bar F_{co_1}^{ij} -( \bar F_{kl}^{on}g_{nm}^{bp}\bar F_{ob}^{ij} + H.\,c.) } \nonumber \\
&&
\stano{- \bar F_{kl}^{oa}g_{am}^{bp}\bar F_{ob}^{ij} -(\bar F_{kl}^{cn}g_{nm}^{bp}\bar F_{cb}^{ij} + H.\,c.)  }
+ (\bar F_{kl}^{ac} \bar g_{am}^{r" p}\bar  F ^{ij}_{r" c} + H.\,c.) 
\lbl{eq:Zijpklm}
\eea

\stano{Note that the formal contractions over the full basis set (\(pq, rs\)) and the complete virtual space (\(ab, cd\)) in the expressions for the \(P_{ij}^{kl}\) and \(Z_{ijp}^{klm}\) matrices [Eqs. (\ref{eq:Pijkl}) and (\ref{eq:Zijpklm})] - as well as the corresponding operations for the $V_{ij}^{kl}$, $X_{ij}^{kl}$, and $B_{ij}^{kl}$ matrices - have been explicitly factorized into individual terms defined over occupied orbital spaces (such as indices \(mn, ob\)). Although this algebraic expansion leads to a higher formal prefactor due to the increased number of individual terms, the overall computational cost is significantly reduced. By avoiding expensive full-space tensor contractions and limiting the scaling primarily to the substantially smaller occupied space, the evaluation of these crucial same-spin triplet components significantly reduces the computational cost.}
\stano{The matrix elements derived above represent the most crucial components for the pseudo-double excitation amplitudes in Eq. (\ref{eq:C2}) within the F12 framework. The remaining terms are significantly smaller \cite{kohn_tew201011} and most of them vanish within the SA.}
\stano{
\bea
\cX_{kl}^{ib} &=&  \bar F _{kl}^{\alpha b} \bar F ^{i}_{\alpha} \\
\cX_{kla}^{ijb} &=& \bar F _{kl}^{\alpha b} \bar F ^{ij}_{\alpha a} + \bar F _{kl}^{cb}(Pi) \bar F ^{ij}_{ca}(Pi) \\
\cW_{kl}^{ij} &=& \half \bar F _{kl}^{\alpha \beta}  F ^{i}_{\alpha}F^{j}_{ \beta} \\
\cJ_{klm}^{ijp} &=& \bar F _{kl}^{\alpha \gamma} \bar g _{\alpha m}^{\beta p}  F ^{i}_{\beta}F^{j}_{ \gamma} \\
\cI_{kl}^{ij} &=&\half \bar F _{kl}^{\alpha \beta}\bar g_{\alpha \beta}^{\gamma \delta}  F ^{i}_{\gamma}F^{j}_{\delta}
+ \bar F _{kl}^{\alpha b}\bar g_{\alpha b}^{\gamma \delta}  F ^{i}_{\gamma}F^{j}_{\delta}
+ \half \bar F _{kl}^{ab}(Pi) \bar g_{ab}^{\gamma \delta}  F ^{i}_{\gamma}F^{j}_{\delta} \\ 
\cI_{lmn}^{ijk} &=& \bar F _{mn}^{\alpha \beta}\bar g_{l \beta}^{\gamma \delta} F ^{i}_{\gamma}F^{j}_{\delta} F^k_{\alpha}
+ \bar F _{mn}^{\alpha b}\bar g_{lb}^{\gamma \delta} F ^{i}_{\gamma}F^{j}_{\delta} F^k_{\alpha} \\
\cZ^{i_1j_1i_2j_2}_{kmnl} &=& \bar F_{kl}^{\alpha\gamma} \bar g_{mn}^{\beta\delta}\bar  F ^{i_1j_1}_{\alpha\beta} \bar  F ^{i_2j_2}_{\delta\gamma}  
+\bar F_{kl}^{a\gamma} \bar g_{mn}^{\beta\delta}\bar  F ^{i_1j_1}_{a\beta} \bar  F ^{i_2j_2}_{\delta\gamma}   
+\bar F_{kl}^{ac}(Pi) \bar g_{mn}^{\beta\delta}\bar  F ^{i_1j_1}_{a\beta} \bar  F ^{i_2j_2}_{\delta c}  \nonumber \\  
&&+\bar F_{kl}^{a\gamma} \bar g_{mn}^{b\delta}\bar  F ^{i_1j_1}_{ab}(Pi) \bar  F ^{i_2j_2}_{\delta\gamma}  
+\bar F_{kl}^{ac}(Pi) \bar g_{mn}^{b\delta} \bar  F ^{i_1j_1}_{ab}(Pi) \bar  F ^{i_2j_2}_{\delta c}  \nonumber \\  
&&+\bar F_{kl}^{ac}(Pi) \bar g_{mn}^{bd}\bar  F ^{i_1j_1}_{ab}(Pi) \bar  F ^{i_2j_2}_{dc}(Pi)   
+\bar F_{kl}^{a\gamma} \bar g_{mn}^{\beta d}\bar  F ^{i_1j_1}_{a\beta} \bar  F ^{i_2j_2}_{d\gamma} 
+\bar F_{kl}^{a\gamma} \bar g_{mn}^{bd}\bar  F ^{i_1j_1}_{ab}(Pi) \bar  F ^{i_2j_2}_{d\gamma} \nonumber \\ 
&&+\bar F_{kl}^{\alpha c} \bar g_{mn}^{\beta\delta}\bar  F ^{i_1j_1}_{\alpha\beta} \bar  F ^{i_2j_2}_{\delta c} 
+\bar F_{kl}^{\alpha c} \bar g_{mn}^{\beta d}\bar  F ^{i_1j_1}_{\alpha\beta} \bar  F ^{i_2j_2}_{dc}(Pi) 
+\bar F_{kl}^{\alpha c} \bar g_{mn}^{b\delta}\bar  F ^{i_1j_1}_{\alpha b} \bar  F ^{i_2j_2}_{\delta c} \\
\cK_{klm}^{aip} &=& \bar F _{kl}^{a\beta} \bar g _{\beta m}^{\gamma p}  F ^{i}_{\gamma}
+ \bar F _{kl}^{ab}(Pi) \bar g _{bm}^{\gamma p}  F ^{i}_{\gamma} \\
\cK^{ijab}_{kmnl} &=& \bar F_{kl}^{\alpha b} \bar g_{mn}^{\beta a}\bar  F ^{ij}_{\alpha\beta} 
+\bar F_{kl}^{\alpha b} \bar g_{mn}^{ca}\bar  F ^{ij}_{\alpha c} 
+\bar F_{kl}^{ab}(Pi) \bar g_{mn}^{ca}\bar  F ^{ij}_{ac}(Pi) \\ 
\cM_{klm}^{aij} &=& \bar F _{kl}^{a \beta} \bar g _{\beta m}^{\gamma \delta}  F ^{i}_{\gamma}F^{j}_{ \delta} 
+ \bar F _{kl}^{ab} \bar g _{b m}^{\gamma \delta}  F ^{i}_{\gamma}F^{j}_{ \delta} \\
\cM_{kmnl}^{aj_1 i_2 j_2} &=& \bar F _{kl}^{a \beta} \bar g _{mn}^{\gamma \delta}  F ^{j_1}_{\gamma} \bar F^{i_2 j_2}_{ \delta \beta}
+ \bar F _{kl}^{a \beta} \bar g _{mn}^{\gamma d}  F ^{j_1}_{\gamma} \bar F^{i_2 j_2}_{d \beta}
+ \bar F _{kl}^{ab}(Pi) \bar g _{mn}^{\gamma d}  F ^{j_1}_{\gamma} \bar F^{i_2 j_2}_{db}(Pi) \\
\cJ_{kmnl}^{i_1 j_1 i_2 j_2} &=& \bar F _{kl}^{\alpha \beta} \bar g _{mn}^{\gamma \delta} F ^{i_1}_{\alpha} F ^{j_1}_{\gamma} \bar F^{i_2 j_2}_{ \delta \beta}
+ \bar F _{kl}^{\alpha b} \bar g _{mn}^{\gamma \delta} F ^{i_1}_{\alpha} F ^{j_1}_{\gamma} \bar F^{i_2 j_2}_{ \delta b}
+ \bar F _{kl}^{\alpha \beta} \bar g _{mn}^{\gamma d} F ^{i_1}_{\alpha} F ^{j_1}_{\gamma} \bar F^{i_2 j_2}_{d \beta} \nonumber \\
&&+ \bar F _{kl}^{\alpha b} \bar g _{mn}^{\gamma d} F ^{i_1}_{\alpha} F ^{j_1}_{\gamma} \bar F^{i_2 j_2}_{db}(Pi)
\eea
}
\subsubsection{Non-iterative perturbative correction of triple excitations}
\label{subsub:triples}

Among the conventional single-reference CC theories, CCSD(T)\cite{ccsdtparent1} has established a status as the "gold standard" of electronic structure theories and has been used as a benchmark for evaluating the performance of diverse wavefunction and/or density functional approaches.  Analogously to CCSD(T)-F12\cite{noga_ten-no2008}, "CCSD0(T)-F12Pi" is formulated here. 
Triple-excitation amplitudes are determined by solving
\be 
\bra {\Phi_0} \tilde a^{ijk}_{abc}( \hat H_1  \hat T_3 + \hat H_2 \hat T_2 +\hat H_2 \hat \cR_2 ) \ket {\Phi_0} = 0 , \lbl{eq:T3F12} 
\ee
more explicitly expressed as 
\bea
&&t^{ijk}_{abc} (f_i^i+f_j^j+f_k^k-f_a^a-f_b^b-f_c^c) = \nonumber \\
&&(\bar g_{de}^{gm}t_{fg}^{no} -\bar g_{ld}^{mn}t_{ef}^{ol} 
 + \half \cV^{m_1n_1o}_{def}c_{m_1n_1}^{mn} )\delta ^{ijk}_{mno}\delta ^{def}_{abc}\,.
\lbl{eq:T3}
\eea
with 
$ \cV^{m_1n_1o}_{def} $ 
defined in Eq. (\ref{eq:V3}).

\subsubsection{Energy within the SP ansatz}

\jozef{Within the standard SP ansatz framework, the universal F12 pseudo-excitation amplitudes are not optimized iteratively but are structurally predefined by the electronic cusp conditions.\cite{Ten_no2004,AdlerWHJ2007,bokhan2008,TEW200832} 
Specifically, the fixed geminal coefficients are uniquely parameterized to satisfy Kato's coalescence constraints across individual spin channels, 
in our case enforcing the universal values of 
\bea
c_{kl}^{ij} = 3/8 \delta_k^i\delta_l^j + 1/8 \delta_l^i\delta_k^j && \mathrm{for\ singlet\ pairs,} \\
c_{kl}^{ij} = 1/4(\delta_k^i\delta_l^j-\delta_l^i\delta_k^j)  && \mathrm{for\ triplet\ pairs.} \eea
} 
Consequently, the energy corresponds to the value of the energy functional with fixed amplitudes of $\hat \cR$ as
\bea E &=& \bra {\Phi}_0 \hat H_N(\hat T_2 + \hat \cR+\half \hat T_1^2) \ket {\Phi}_0 _{\mathrm C} \nonumber \\
&& + \bra {\Phi}_0 \hat \cR^{\dagger} \hat H_N (1+\hat \cR + \hat  T_2 + \hat \cR \hat T_2 ) \ket {\Phi}_0 _{\mathrm C} \nonumber \\
&& + \bra {\Phi}_0 \hat \cR^{\dagger} \hat H_N ( \half {\hat \cR}^2 +\hat \cR \hat T_1+\half \hat T_1^2 + \half \hat \cR \hat T_1^2 ) \ket {\Phi}_0 _{\mathrm C} . \lbl{eq:lagr} 
\eea
As compared to the first two, contribution from the last term is negligible and can be safely neglected.
Nevertheless, 
conventional $t$ amplitudes are updated  
in each iteration as described in the previous subsections. 
Having this in mind, a constant term can be identified in Eq. (\ref{eq:lagr}), namely:
\be
E_{\cR} =  (2 \cV_{ij}^{kl} 
+ \cB_{ij}^{kl}+\tilde \cP_{ij}^{kl} + \cX_{ij}^{kl}(\epsilon_i+\epsilon_j) 
+ \cZ_{imj}^{klm} + g^{mn}_{ij} \cX_{mn}^{kl}) c_{kl}^{ij}\,, 
\ee 
where 
a relation $\cV_{ij}^{kl} c_{kl}^{ij} = c^{kl}_{ij} (\cV^\dagger)^{ij}_{kl} $
is used.

An explicit formula for the energy can then be written
as
\be 
E = (g_{ij}^{ab}+ \half (\cV^\dagger)_{kl}^{ab} c^{kl}_{ij}) (\half t^{ij}_{ab} + t^i_a t^j_b) + (\cV^\dagger)_{ij}^{aj} t^i_a + E_{\cR}.
\lbl{eq:EnExpl}
\ee

\subsection{Cluster-expansion approach to orbital theories}
\label{subsec:cluster-expansion}
\stanoo{Following the pioneering work of Hirao and Nakatsuji\cite{hirao-closed}, we summarize and recapitulate the cluster-expansion analysis of closed-shell orbital theories in Sections \ref{subsubsec:cluster-expansion1} through \ref{subsubsec:cluster-expansion3} to establish the theoretical foundation for our regularized framework.} 

\subsubsection{Transformation of a Slater determinant with spin-adapted cluster operators}
\label{subsubsec:cluster-expansion1}

Consider optimizing a wavefunction of a closed-shell system expressed by a single Slater determinant $ |\Phi_0\rangle $ 
within the orbital-theory framework. Transformation of a Slater determinant to another is expressed using a single-excitation operator
\begin{equation}
    \hat{T}_1 = t^{i\sigma}_{a\sigma\prime} a^{\dagger}_{a\sigma\prime} a_{i\sigma}
\end{equation}
in two ways; \jozef{using} the Thouless theorem\cite{thouless}
\begin{equation}
    |\Phi^{\prime}\rangle = \mathcal{N}\exp{(\hat{T}_1)}|\Phi_0\rangle,
\end{equation}
where $\mathcal{N}$ is a normalization factor, or \jozef{via} unitary transformation
\begin{equation}
    |\Phi^{\prime}\rangle = \exp{(\hat{T}_1-\hat{T}_1^{\dagger})}|\Phi_0\rangle.
\end{equation}
The transformed Slater determinants given in the two ways are equivalent.\cite{hirao-closed} The single-excitation operator $\hat{T}_1$ can be decomposed into a spin-adapted form
\begin{eqnarray}
    \hat{T}_1 = w^{\prime i}_{(0)a}\tilde{S}^{(0)a\dagger}_i &&+ \sum_{\mu=x,y,z}w^{\prime i}_{(\mu)a}\tilde{S}^{(\mu)a\dagger}_i,    \\
    \tilde{S}^{(0)a\dagger}_i &&= \frac{1}{\sqrt{2}}\mathbf{a_{\mathbf{a}}^{\dagger}}\sigma_0\mathbf{a_{\mathbf{i}}},    \\
    \tilde{S}^{(\mu)a\dagger}_i &&= \frac{1}{\sqrt{2}}\mathbf{a_{\mathbf{a}}^{\dagger}}\sigma_{\mu}\mathbf{a_{\mathbf{i}}} \quad (\mu=x,y,z),
\end{eqnarray}
where 
\begin{equation}
    \mathbf{a_{\mathbf{a}}^{\dagger}} = (a_{a\alpha}^{\dagger}, a_{a\beta}^{\dagger}), \mathbf{a_{\mathbf{i}}} = \begin{pmatrix} a_{i\alpha} \\ a_{i\beta} \end{pmatrix}
\end{equation}
and $\sigma_{\mu} (\mu = 0,1,2,3)$ are the Pauli matrices. The $w^{\prime i}_{(\mu)a}$ amplitudes are complex numbers 
($w^{\prime i}_{(\mu)a} \in \mathbb C$)
and expressed by $w^{\prime} = u^{\prime} + iv^{\prime}$ using real numbers $u^{\prime}$ and $v^{\prime}$ 
($u^{\prime}, v^{\prime} \in \mathbb R$).
The $\tilde{S}^{(0)a\dagger}_i$ operators generate spin-singlet configurations from a closed-shell one like
\begin{equation}
    \tilde{S}^{(0)a\dagger}_i|\Phi_0\rangle = \frac{1}{\sqrt{2}}[\det(\phi_1\alpha\phi_1\beta\cdots\phi_i\alpha\phi_a\beta\cdots\phi_n\alpha\phi_n\beta) - \det(\phi_1\alpha\phi_1\beta\cdots\phi_i\beta\phi_a\alpha\cdots\phi_n\alpha\phi_n\beta)],
\end{equation}
while $\tilde{S}^{(\bar{k})a\dagger}_i$ generate spin-triplet configurations
\begin{eqnarray}
	\frac{1}{\sqrt{2}}(\tilde{S}^{(x)a\dagger}_i+i\tilde{S}^{(y)a\dagger}_i)|\Phi_0\rangle &=& \det(\phi_1\alpha\phi_1\beta\cdots\phi_i\alpha\phi_a\alpha\cdots\phi_n\alpha\phi_n\beta),    \\
	\tilde{S}^{(z)a\dagger}_i|\Phi_0\rangle &=& \frac{1}{\sqrt{2}}[\det(\phi_1\alpha\phi_1\beta\cdots\phi_i\alpha\phi_a\beta\cdots\phi_n\alpha\phi_n\beta) \nonumber \\ 
	&& + \det(\phi_1\alpha\phi_1\beta\cdots\phi_i\beta\phi_a\alpha\cdots\phi_n\alpha\phi_n\beta)],    \\
	\frac{1}{\sqrt{2}}(\tilde{S}^{(x)a\dagger}_i-i\tilde{S}^{(y)a\dagger}_i)|\Phi_0\rangle &=& \det(\phi_1\alpha\phi_1\beta\cdots\phi_i\beta\phi_a\beta\cdots\phi_n\alpha\phi_n\beta).
\end{eqnarray}
The Hermitian conjugate of $\tilde{S}^{\dagger}$ is defined by $\tilde{S} \equiv (\tilde{S}^{\dagger})^{\dagger}$. The $\tilde{S}$ and $\tilde{S}^{\dagger}$ operators satisfy quasi-boson commutation relations as follows:
\begin{eqnarray}
    [\tilde{S}^{(\mu)a}_i,\tilde{S}^{(\nu)b}_j] = [\tilde{S}^{(\mu)a\dagger}_i,\tilde{S}^{(\nu)b\dagger}_j] = 0,    \\
    \langle\Phi_0|[\tilde{S}^{(\mu)a}_i,\tilde{S}^{(\nu)b\dagger}_j]|\Phi_0\rangle = \delta_{\mu\nu}\delta_{ij}\delta_{ab}.
\end{eqnarray}
Because these quasi-commutation relations hold for every $\mu = 0,1,2,3$, the four species of the spin-adapted excitation operators are independent of each other.

\subsubsection{Stability and symmetry dilemmas solved by projection}
\label{subsubsec:cluster-expansion2}

Consider seeking the solution with the lowest energy eigenvalue of the Schr\"{o}dinger equation of a closed-shell system with the Hartree--Fock (HF) theory. The HF ansatz is obtained by rotating a Slater determinant in the space spanned by determinants generated by the real spin-singlet excitation operators, i.e.
\begin{equation}
    |\Phi_{\mathrm{HF}}\rangle = \mathcal{N}\exp{(u^{\prime i}_{(0)a}\tilde{S}^{(0)a\dagger}_i)}|\Phi_0\rangle,
\end{equation}
or
\begin{equation}
    |\Phi_{\mathrm{HF}}\rangle = \exp{[u^{\prime i}_{(0)a}(\tilde{S}^{(0)a\dagger}_i-\tilde{S}^{(0)a}_i)}]|\Phi_0\rangle.
\end{equation}
The variational space can be extended within the orbital-theory framework by making the amplitudes complex like $u^{\prime} \to w^{\prime} = u^{\prime} + iv^{\prime}$ where
$u^{\prime}, v^{\prime} \in \mathbb R$
and adding the spin-triplet operators $\tilde{S}^{(\bar{k})\dagger} (\bar{k} = x,y,z)$ to the exponent. The variational principle indicates that such an extension of the variational space gives an energy lower than that obtained with the HF theory. However, if the HF ansatz is stable, energy closer to the exact one cannot be obtained by extending the variational space within the orbital-theory framework. Hirao and Nakatsuji introduced the concept named "stability dilemma" indicating this situation\cite{hirao-closed}. Meanwhile, an energy lower than that of RHF is obtained in cases where the RHF solution is unstable. In these cases, the wavefunction does not hold the symmetry imposed by the Hamiltonian, 
e.g.
although UHF describes qualitatively correct dissociation curves, the UHF wavefunctions are not eigenfunctions of the spin operator $\mathbf{S}^2$ (symmetry dilemma).\cite{lykos}

For solving the stability and symmetry dilemmas, Hirao and Nakatsuji classified the HF stability conditions by adapting the spin symmetry to the variational space. The \stanoo{quadratic} term of the energy in the expansion around the HF energy $E_{\mathrm{HF}}$ by deviation of the wavefunction from $|\Phi_{\mathrm{HF}}\rangle$ is given in the spin-adapted form,
\begin{equation}
    \frac{1}{2} \begin{pmatrix} \mathbf{w}^{\prime}_{(0)} \\ \mathbf{w}^{\prime\ast}_{(0)} \end{pmatrix}^{\dagger} \begin{pmatrix} \mathbf{A}_{(0)} & \mathbf{B}_{(0)} \\ \mathbf{B}_{(0)}^{\ast} & \mathbf{A}_{(0)}^{\ast} \end{pmatrix} \begin{pmatrix} \mathbf{w}^{\prime}_{(0)} \\ \mathbf{w}^{\prime\ast}_{(0)} \end{pmatrix} + \frac{1}{2} \sum_{\bar{k}=x,y,z} \begin{pmatrix} \mathbf{w}^{\prime}_{(\bar{k})} \\ \mathbf{w}^{\prime\ast}_{(\bar{k})} \end{pmatrix}^{\dagger} \begin{pmatrix} \mathbf{A}_{(\bar{k})} & \mathbf{B}_{(\bar{k})} \\ \mathbf{B}_{(\bar{k})}^{\ast} & \mathbf{A}_{(\bar{k})}^{\ast} \end{pmatrix} \begin{pmatrix} \mathbf{w}^{\prime}_{(\bar{k})} \\ \mathbf{w}^{\prime\ast}_{(\bar{k})} \end{pmatrix} \ge 0, \label{eq:HFstability}
\end{equation}
where the submatrices are defined by
\begin{eqnarray}
    (\mathbf{A}_{(0)})^{ab}_{ij} = \langle\Phi_{\mathrm{HF}}|\tilde{S}^{(0)a}_i\hat{H}\tilde{S}^{(0)b\dagger}_j-E_{\mathrm{HF}}|\Phi_{\mathrm{HF}}\rangle, \\
    (\mathbf{B}_{(0)})^{ab}_{ij} = \langle\Phi_{\mathrm{HF}}|\tilde{S}^{(0)a}_i\tilde{S}^{(0)b}_j\hat{H}|\Phi_{\mathrm{HF}}\rangle, \\
    (\mathbf{A}_{(\bar{k})})^{ab}_{ij} = \langle\Phi_{\mathrm{HF}}|\tilde{S}^{(\bar{k})a}_i\hat{H}\tilde{S}^{(\bar{k})b\dagger}_j-E_{\mathrm{HF}}|\Phi_{\mathrm{HF}}\rangle, \\
    (\mathbf{B}_{(\bar{k})})^{ab}_{ij} = \langle\Phi_{\mathrm{HF}}|\tilde{S}^{(\bar{k})a}_i\tilde{S}^{(\bar{k})b}_j\hat{H}|\Phi_{\mathrm{HF}}\rangle \quad (\bar{k}=x,y,z).
\end{eqnarray}
Under an approximation relying on the sparsity of the submatrices $\mathbf{A}$ and $\mathbf{B}$, the four species of the HF stability condition are obtained as follows,
\begin{eqnarray}
	E^{(0)aa}_{ii}-E_{\mathrm{HF}}+K^{aa\ast}_{ii} \ge 0 \quad (\mathbf{A}_{(0)}+\mathbf{B}_{(0)} \ge 0, \quad \mathrm{real} \ \mathrm{spin-singlet} \ \mathrm{stability} \ \mathrm{condition}),&& \label{eq:real-singlet} \\
	E^{(0)aa}_{ii}-E_{\mathrm{HF}}-K^{aa\ast}_{ii} \ge 0 \quad (\mathbf{A}_{(0)}-\mathbf{B}_{(0)} \ge 0, \quad \mathrm{imaginary} \ \mathrm{spin-singlet} \ \mathrm{stability} \ \mathrm{condition}),&& \label{eq:imaginary-singlet} \\
	E^{(\bar{k})aa}_{ii}-E_{\mathrm{HF}}-K^{aa\ast}_{ii} \ge 0 \quad (\mathbf{A}_{(\bar{k})}+\mathbf{B}_{(\bar{k})} \ge 0, \quad \mathrm{real} \ \mathrm{spin-triplet} \ \mathrm{stability} \ \mathrm{condition}),&& \label{eq:real-triplet} \\
	E^{(\bar{k})aa}_{ii}-E_{\mathrm{HF}}+K^{aa\ast}_{ii} \ge 0 \quad (\mathbf{A}_{(\bar{k})}-\mathbf{B}_{(\bar{k})} \ge 0, \quad \mathrm{imaginary} \ \mathrm{spin-triplet} \ \mathrm{stability} \ \mathrm{condition}),&& \label{eq:imaginary-triplet}
\end{eqnarray}
where 
\begin{eqnarray}
    E^{(0)aa}_{ii} &&= \langle\Phi_{\mathrm{HF}}|\tilde{S}^{(0)a}_i\hat{H}\tilde{S}^{(0)a\dagger}_i|\Phi_{\mathrm{HF}}\rangle, \\
    E^{(\bar{k})aa}_{ii} &&= \langle\Phi_{\mathrm{HF}}|\tilde{S}^{(\bar{k})a}_i\hat{H}\tilde{S}^{(\bar{k})a\dagger}_i|\Phi_{\mathrm{HF}}\rangle, \\
    K^{aa\ast}_{ii} &&= \langle\Phi_{\mathrm{HF}}|\tilde{S}^{(0)a}_i\tilde{S}^{(0)a}_i\hat{H}|\Phi_{\mathrm{HF}}\rangle = -\langle\Phi_{\mathrm{HF}}|\tilde{S}^{(\bar{k})a}_i\tilde{S}^{(\bar{k})a}_i\hat{H}|\Phi_{\mathrm{HF}}\rangle.
\end{eqnarray}

It should be emphasized that $K^{aa\ast}_{ii}$ is the exchange integral. 

\stano{Crucially, Eqs. (\ref{eq:imaginary-singlet}) and (\ref{eq:real-triplet}) demonstrate that the onset conditions for \stanoo{both} the imaginary spin-singlet instability and the real spin-triplet instability 
\stanoo{are fundamentally driven by the negative exchange integral $-K^{aa\ast}_{ii}$. 
This shared exchange-driven mechanism distinguishes them from Eqs. (\ref{eq:real-singlet}) and (\ref{eq:imaginary-triplet}), providing a sound physical basis}
for bridging the gap between the complex molecular orbital (CMO) framework \cite{MAYER1980} and the real triplet instabilities governing the homolytic cleavage of \(\text{N}_{2}\) and \(\text{CO}_{2}\), rendering the \(\text{CCSD0}\) amplitude regularization directly relevant to the description of bond-breaking regions.}

Now look at the real spin-triplet stability condition
(\ref{eq:real-triplet}).
As explained earlier, an orbital theory with an energy lower than RHF is not obtained by extending the variational space if the stability condition is met (stability dilemma), while it is if the RHF wavefunction is unstable although the obtained UHF wavefunction does not preserve the spin symmetry (symmetry dilemma). Apply the spin projection operator
\begin{equation}
    \mathcal{O}_S = \prod_{l \ne S} \frac{\mathbf{S}^2-l(l+1)\mathbf{1}}{S(S+1)-l(l+1)},
\end{equation}
proposed by L\"{o}wdin\cite{lowdin1955,lowdin1963,pauncz1979,pauncz2000} with $(S=0)$ to the wavefunction $|\Phi_{\mathrm{HF}}\rangle$ to extract the spin-singlet component. This projection erases the $\tilde{S}^{(\bar{k})a\dagger}_i|\Phi_{\mathrm{HF}}\rangle$ configurations because they are of pure spin-triplet. As a consequence, the $\mathbf{A}_{(\bar{k})}$ submatrices, which derive from the single-excitation configurations, vanish from the stability condition 
(\ref{eq:HFstability}).
The new stability condition $(\mathbf{B}_{(\bar{k})} \ge 0)$ is not satisfied because the trace of $\mathbf{B}_{(\bar{k})}$ is not positive,
\begin{equation}
    \mathrm{Tr}(\mathbf{B}_{(\bar{k})}) = \sum_{ia} \langle\Phi_{\mathrm{HF}}|\tilde{S}^{(\bar{k})a}_i\tilde{S}^{(\bar{k})a}_i\hat{H}|\Phi_{\mathrm{HF}}\rangle = -\sum_{ia} K^{aa\ast}_{ii} \le 0,
\end{equation}
and thus $\mathbf{B}_{(\bar{k})}$ has at least one negative eigenvalue. Now an orbital theory, which solves both the real spin-triplet stability and symmetry dilemmas, is obtained by projection of UHF to the spin-singlet state. The obtained orbital theory is called "extended Hartree--Fock (EHF) theory" in Mayer's nomenclature\cite{MAYER1980} and the ansatz is expressed using the spin-adapted cluster operators as follows:
\begin{equation}
    |\Phi_{\mathrm{EHF}}\rangle = \mathcal{N}\mathcal{O}_S \exp{(u^{\prime i}_{(0)a}\tilde{S}^{(0)a\dagger}_i + u^{\prime i}_{(z)a}\tilde{S}^{(z)a\dagger}_i)} |\Phi_0\rangle |_{S=0}.
\end{equation}
In the same manner, the imaginary spin-singlet stability and symmetry dilemmas are solved by the projection operator $\mathcal{O}_R$, which extracts the real part of a function, i.e. $\mathcal{O}_R\Psi = \Re\Psi$. 
The obtained complex molecular orbital theory in Hendekovi\'{c}'s nomenclature\cite{hendekovic} and the ansatz is expressed using the spin-adapted cluster operators as follows:
\begin{equation}
    |\Phi_{\mathrm{CMO}}\rangle = \mathcal{N}\mathcal{O}_R \exp{(w^{\prime i}_{(0)a}\tilde{S}^{(0)a\dagger}_i)} |\Phi_0\rangle.
\end{equation}
Contrary to the real spin-triplet and imaginary spin-singlet cases, the stability and symmetry dilemmas are not solved in the imaginary spin-triplet and real spin-singlet cases within the orbital theory framework. 

\subsubsection{Configuration analysis for projected HF theories}
\label{subsubsec:cluster-expansion3}

The EHF and CMO wavefunctions can be expanded by the reference Slater determinants consisting of the natural orbitals (NOs), which can be approximated as that of RHF. In the case of the UHF wavefunction of a $2n$-electron system, which gives an EHF wavefunction when the projection operator $\mathcal{O}_S$ is applied, the NOs $(\lambda_k, \nu_k)$ come from the orbitals with $\alpha$ and $\beta$ spins,\jozef{ $(k = 1 \cdots n)$,}
\begin{eqnarray}
    \phi^{\mathrm{UHF}}_{k\alpha}(\mathbf{r}) &&= \xi_k \lambda_k(\mathbf{r}) + \eta_k \nu_k(\mathbf{r}), \\
    \phi^{\mathrm{UHF}}_{k\beta}(\mathbf{r}) &&= \xi_k \lambda_k(\mathbf{r}) - \eta_k \nu_k(\mathbf{r}) ,
\end{eqnarray}
\jozef{where $\xi_k$ and $\eta_k$ represent the real normalization coefficients (orbital occupation amplitudes) that strictly satisfy the condition $\xi_k^2 + \eta_k^2 = 1$. In closed-shell equilibrium structures, $\xi_k \approx 1$ and $\eta_k \approx 0$, whereas in strongly correlated regions where the RHF solution undergoes homolytic molecular orbital breakdown, the increasing magnitude of $\eta_k$ directly quantifies the emerging multireference character of the system.}

Since the $\eta_k$ coefficients are sufficiently small, the UHF wavefunction can be expanded by a reference Slater determinant consisting of $\lambda_k(\mathbf{r})$,
\begin{eqnarray}
    |\Phi_{\mathrm{UHF}}\rangle &&= C_0 |\Phi_{\mathrm{ref}}\rangle + \sum_k C_{\mathrm{S}}(kk^{\ast})|\Phi_{\mathrm{S}}(kk^{\ast})\rangle \nonumber \\ &&+ \sum_k C_{\mathrm{D}}(kk^{\ast})|\Phi_{\mathrm{D}}(kk^{\ast})\rangle + \sum_{k<l} C_{\mathrm{D}}(kk^{\ast};ll^{\ast})|\Phi_{\mathrm{D}}(kk^{\ast};ll^{\ast})\rangle + \cdots, \\
    |\Phi_{\mathrm{ref}}\rangle &&= \det(\lambda_1 \alpha \lambda_1 \beta \cdots \lambda_k \alpha \lambda_k \beta \cdots \lambda_n \alpha \lambda_n \beta), \label{eq:refUHF} \\
    |\Phi_{\mathrm{S}}(kk^{\ast})\rangle &&= \det[\lambda_1 \alpha \lambda_1 \beta \cdots \lambda_k \nu_k \Big(\frac{\alpha\beta+\beta\alpha}{\sqrt{2}}\Big) \cdots \lambda_n \alpha \lambda_n \beta], \\
	|\Phi_{\mathrm{D}}(kk^{\ast})\rangle &&= \det(\lambda_1 \alpha \lambda_1 \beta \cdots \nu_k \alpha \nu_k \beta \cdots \lambda_n \alpha \lambda_n \beta), \label{eq:DexP} \\
    |\Phi_{\mathrm{D}}(kk^{\ast};ll^{\ast})\rangle &&= \det[\lambda_1 \alpha \lambda_1 \beta \cdots \lambda_k \nu_k \lambda_l \nu_l \Big(\frac{(\alpha\beta+\beta\alpha)(\alpha\beta+\beta\alpha)}{2}\Big) \cdots \lambda_n \alpha \lambda_n \beta].
\end{eqnarray}

\jozef{Here, the subscripts S and D explicitly denote single- and double-excitation configurations, respectively, generated relative to the closed-shell reference state.}
The expansion of the EHF wavefunction by $|\Phi_{\mathrm{ref}}\rangle$ is obtained by applying $\mathcal{O}_S$ to $|\Phi_{\mathrm{UHF}}\rangle$. The $|\Phi_{\mathrm{S}}(kk^{\ast})\rangle$ determinants vanish because they are of pure spin-triplet state. While $|\Phi_{\mathrm{D}}(kk^{\ast})\rangle$ remains unchanged, the 
unpaired
double-excitation determinants $|\Phi_{\mathrm{D}}(kk^{\ast};ll^{\ast})\rangle$ are affected by the projection, 
\begin{eqnarray}
    &&|\Phi_{\mathrm{EHF}}\rangle \nonumber \\ &&= C_0 |\Phi_{\mathrm{ref}}\rangle + \sum_k C_{\mathrm{D}}(kk^{\ast})|\Phi_{\mathrm{D}}(kk^{\ast})\rangle + \sum_{k<l} C_{\mathrm{D}}(kk^{\ast};ll^{\ast})\mathcal{O}_S|\Phi_{\mathrm{D}}(kk^{\ast};ll^{\ast})\rangle |_{S=0} + \cdots, \label{eq:EHF} \\
    &&\mathcal{O}_S|\Phi_{\mathrm{D}}(kk^{\ast};ll^{\ast})\rangle|_{S=0}=  \det[\lambda_1 \alpha \lambda_1 \beta \nonumber \\ &&\cdots \lambda_k \nu_k \lambda_l \nu_l \Big(\frac{(2\alpha\alpha\beta-\alpha\beta\alpha-\beta\alpha\alpha)\beta-(\alpha\beta\beta+\beta\alpha\beta-2\beta\beta\alpha)\alpha}{\sqrt{12}}\Big) \cdots \lambda_n \alpha \lambda_n \beta].
\end{eqnarray}
The CMO wavefunction is similarly expanded: 
\begin{eqnarray}
    |\Phi_{\mathrm{CMO}}\rangle = C_0 |\Phi_{\mathrm{ref}}\rangle + \sum_k C_{\mathrm{D}}(kk^{\ast})|\Phi_{\mathrm{D}}(kk^{\ast})\rangle + \sum_{k<l} C_{\mathrm{D}}(kk^{\ast};ll^{\ast})\mathcal{O}_R|\Phi_{\mathrm{D}}(kk^{\ast};ll^{\ast})\rangle + \cdots, \label{eq:CMO} 
\end{eqnarray}
\jozef{with $|\Phi_{\mathrm{ref}}\rangle$ and $|\Phi_{\mathrm{D}}(kk^{\ast})\rangle$ formally defined as in Eqs. (\ref{eq:refUHF}) and (\ref{eq:DexP}), respectively. The projection operator $\mathcal{O}_{R}$, which extracts the real part of the wavefunction, acts on the unpaired double-electronic configurations as follows:}
\begin{eqnarray}
    \mathcal{O}_R|\Phi_{\mathrm{D}}(kk^{\ast};ll^{\ast})\rangle = \det[\lambda_1 \alpha \lambda_1 \beta \cdots \lambda_k \nu_k \lambda_l \nu_l \Big(\frac{(\alpha\beta-\beta\alpha)(\alpha\beta-\beta\alpha)}{2}\Big) \cdots \lambda_n \alpha \lambda_n \beta].
\end{eqnarray}
As Mayer showed, EHF qualitatively describes covalent-bond dissociation without spin contamination.\cite{MAYER1980} Nevertheless, the spin part of the leading configurations of EHF are not along the intuitive idea of the covalent-bond dissociation. Meanwhile, the leading configurations of CMO are similar to those of EHF except for the 
unpaired
double-electronic excitations and rather along with the intuitive picture of the dissociation. This mathematical analysis supports the observation by Small
\textit{et al.}
that cRHF is closely related to UHF and performs well for describing dissociations and transition states.\cite{Small2015}

\subsubsection{Evaluation of CCSD0 with help of the cluster-expansion approach}
\label{subsubsec:cluster-expansion4}

\stanoo{The configuration structures of the projected frameworks can be directly mapped to the regularized coupled-cluster formalisms outlined in Section \ref{subsec:ccsd0-f12pi}. 
In the CCSD0 theory, the doubles-excitation operators are restricted to the spin-singlet pairing space according to Eqs. (\ref{eq:CCSD0_ansatz})–(\ref{eq:Pab_def}).}
The $\exp{(\hat{T}_1)}$ operator just transforms the reference Slater determinant $|\Phi_0\rangle$ as shown by the Thouless theorem.\cite{thouless} The CCSD0 wavefunction is then expanded by 
the Brueckner reference $|\Phi_{\mathrm{B}}\rangle = \exp{(\hat{T}_1)}|\Phi_0\rangle$.
The leading configurations given by 
$(1+\frac{1}{2} \sigma^{ab}_{ij}\hat{P}^{\dagger}_{ab}\hat{P}_{ij})|\Phi_{\mathrm{B}}\rangle$
are identical to those of the CMO wavefunction 
(\ref{eq:CMO})
under an approximation that both the reference Slater determinants can be regarded as the RHF wavefunction. Therefore, 
\stanoo{CCSD0 shares its leading paired configurations with CMO.} 

Since strong correlation is related to the HF instabilities, the correspondence between CCSD0 and CMO indicates the capability of CCSD0 in describing strong correlation. As already shown, the orbital theories which overcome the cases of imaginary spin-singlet and real spin-triplet, CMO and EHF\cite{hirao-closed} (even those before the projections, cRHF and UHF\cite{Small2015}), are closely related. Moreover, the HF stability conditions 
[Eqs. (\ref{eq:real-singlet}) to (\ref{eq:imaginary-triplet})]
indicate that the imaginary spin-singlet instability precedes the real spin-singlet one. They also indicate that the real spin-triplet instability precedes the imaginary spin-triplet one. These facts may indicate that CCSD0 describes strong correlation of closed-shell systems in most of the cases with limited exceptions. 

There are two differences between CCSD0 and CMO, which give CCSD0 an advantage over CMO: First, CCSD0 is size consistent while CMO is not. Second, the CCSD0 wavefunction has $\frac{n(n+1)}{2}$ parameters while the CMO wavefunction has only $n$ parameters. It should also be emphasized that implementation of CCSD0 is more straightforward than that of CMO because all the algorithms of CCSD0 are common to normal CCSD. 
\jozef{Since CMO is an orbital theory formulated based on the mean-field approximation, the common features of
} \shota{
CCSD0 and CMO highlight that the single-reference ansatz remains structurally well-behaved in degenerate regions due to the restriction to the singlet-paired space. 
} 
\stanoo{In summary, while this configuration analysis provides a helpful structural correspondence to the projected orbital theories, the remaining triplet correlation energy, which is neglected due to the omission of the same-spin triplet channels, is subsequently supplemented by our explicitly correlated framework detailed in Section \ref{subsec:ccsd0-f12pi} .}

\section{Results and {D}iscussion}
All computations were carried out by using our local version of the DIRCCR12-OS code.\cite{DIRCCR12} For evaluation of matrix elements specific to explicitly correlated part, approximation C was used.\cite{SKappC05} Frozen-core approximation and Dunning's correlation consistent basis sets\cite{Dunning:89,Dunning:92,Dunning:93,Dunning:94,Dunning:95} were used throughout. 
\stano{The present pilot molecular calculations have been carried out within the framework of the Standard Approximation (SA). Consequently, the single-electron pseudo-excitations are omitted, and the coupling terms involving the complementary auxiliary basis set space (CABS) are neglected rather than rigorously eliminated. While a full CABS treatment would provide additional basis-set flexibility, the current SA implementation is formally sufficient to demonstrate the physical viability of the CCSD0-F12Pi scheme.}

For the nitrogen molecule, the dissociation curve was calculated at the RHF, CCSD, CCSD-F12, CCSD0 and CCSD0-F12Pi levels 
in 
conjunction with cc-pVTZ. For the carbon dioxide molecule, the symmetric dissociation curve was calculated at the RHF, CCSD(T), CCSD(T)-F12, CCSD0(T) and CCSD0(T)-F12Pi levels 
in 
conjunction with cc-pVTZ. For the symmetric linear methylene (CH$_2$) molecule with bond length of 1.07 {\AA}  as well as the cyclobutadiene (C$_4$H$_4$) molecule, the energy was calculated with CCSD0, CCSD0-F12Pi SP, CCSD0-F12Pi (full), and those with perturbative triples, adopting the basis sets cc-pVXZ (X = D, T, Q and 5). The (transition-state) geometry of C$_4$H$_4$ was the same as that used in \shota{Ref.} \citenum{mkcc_our1} : The C--C and C--H lengths were 1.44400 and 1.07600 \AA, respectively. The results for CH$_2$ and C$_4$H$_4$ were compared with the energies calculated with MkCCSD, MkCCSD(T), MkCCSD-F12 and MkCCSD(T)\stanoo{-F12}\cite{MkCCSD,mkcc_our1}, models of Mk-MRCC.\cite{MkMRCC-1,MkMRCC-2,MkMRCC-3,maitra_mukherjee}  
\stanoo{All potential energy curves and spectroscopic constants for $\mathrm{N_2}$ (Figure \ref{fig:N2}, Table \ref{n2spectro_combined}) 
and $\mathrm{CO_2}$ (Figure \ref{fig:CO2}) were calculated using the fully optimized F12 amplitude version of the CCSD0-F12Pi and CCSD0(T)-F12Pi methodologies.}

\subsection{Spectroscopic Analysis and Dissociation of the $\mathrm{N_2}$ Molecule}

The potential energy curve (PEC) for the homonuclear cleavage of the triple bond in $\mathrm{N_2}$ is one of
the most severe among typical cases for testing the tolerance of single-reference electronic structure methodologies due to the rapid emergence of strong static correlation. 

\begin{figure}[h]
\centering
\includegraphics[width=12cm]{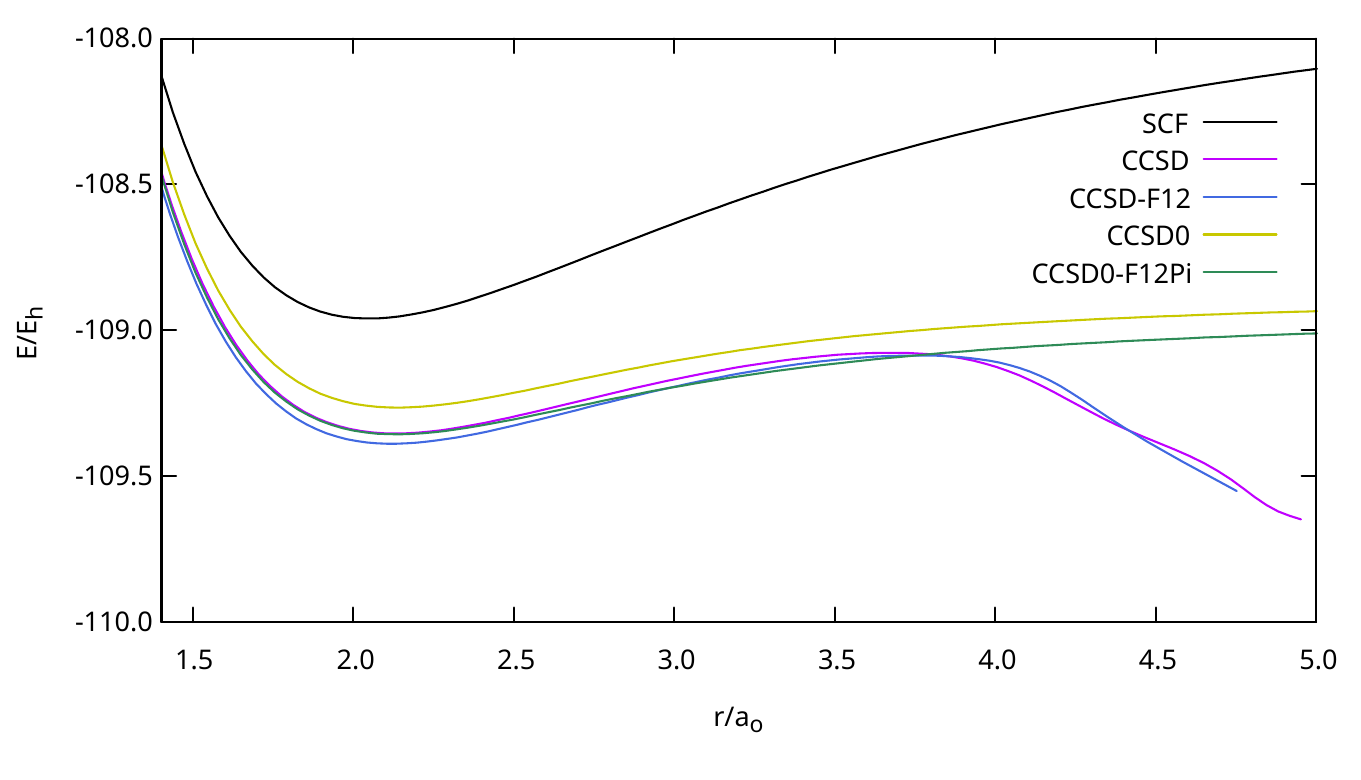}
\caption{\shota{Calculated potential energy curves for $\mathrm{N_2}$ with the cc-pVTZ basis set} as a function of bond length (in the atomic unit) calculated at the (black) RHF, (purple) CCSD, (blue) CCSD-F12, (yellow) CCSD0 and (green) CCSD0-F12Pi levels.}
\label{fig:N2}
\end{figure} 

Figure \ref{fig:N2} illustrates the PECs computed at the RHF, CCSD, CCSD-F12, CCSD0, and the newly developed CCSD0-F12Pi levels of theory using the cc-pVTZ basis set. In the stretched region ($r > 3.5\,a_0$), standard non-variational CCSD and explicitly correlated CCSD-F12 exhibit catastrophic breakdowns, resulting in severe unphysical energy drops 
driven by the singlet-triplet instabilities of the underlying RHF reference. Conversely, the singlet-paired CCSD0 framework effectively regularizes the amplitude equations, securing a qualitatively correct, smooth, and monotonic asymptotic behavior.

Quantitatively, however, the parent CCSD0 model suffers from a significant deficit due to the deliberate omission of the same-spin triplet excitation channels. Our CCSD0-F12Pi ansatz successfully remediates this shortcoming. 
While opposite-spin singlet interactions are treated via standard explicit terms ($\alpha\beta + a\beta$), the same-spin triplet channels are recovered through the newly introduced $a,b$ operator terms within the F12 framework \stano{\mbox{(eq. \ref{eq:pseudo2Pi})}}.
These explicit pseudo-excitations are optimized simultaneously with the cluster amplitudes, \stano{ seamlessly recovering the same-spin correlation effects that are inherently omitted in the conventional CCSD0 framework.}
Near the equilibrium minimum ($r \approx 2.1\,a_0$), the CCSD0-F12Pi curve recovers the lost dynamic correlation, shifting the energy downward by approximately $0.1\,E_h$ and running perfectly parallel to the high-level conventional CCSD baseline before seamlessly transitioning into a stable dissociation path.

\begin{table}[h]
\caption{Spectroscopic constants and dissociation energies of $\mathrm{N_2}$ computed with the CCSD0(T)-F12Pi method. Comparison of the Dunham polynomial expansion across cc-pVTZ and cc-pV5Z basis sets, alongside the global Morse potential fit \stanoo{across cc-pV5Z basis set} against experimental reference values.\label{n2spectro_combined}}
\centering
\footnotesize
\begin{tabular}{lcccccc}
\tableline
& cc-pVTZ Basis Set && \multicolumn{2}{c}{cc-pV5Z Basis Set} & & \\
\cline{2-2} \cline{4-5}
Spectroscopic Constant & Dunham Fit && Dunham Fit & Morse Fit & Experiment & Unit \\
\tableline
$R_e$                  & 1.1056      && 1.1061      & 1.1084      & 1.0977      & \AA \\
$V_e$ (minimum energy) & -109.377301 && -109.381860 & -109.382381 & —           & $E_\mathrm{h}$ \\
$\omega_e$             & 2277.9      && 2262.0      & 2247.9      & 2358.6      & $\mathrm{cm}^{-1}$ \\
$\omega_e x_e$         & 15.84       && 15.87       & 15.54       & 14.32       & $\mathrm{cm}^{-1}$ \\
$B_e$                  & 1.969       && 1.968       & 1.960       & 1.998       & $\mathrm{cm}^{-1}$ \\
$\alpha_e$             & 0.0200      && 0.0200      & 0.0185      & 0.0173      & $\mathrm{cm}^{-1}$ \\
ZPE                    & 0.00517     && 0.00514     & 0.00510     & 0.00537     & $E_\mathrm{h}$ \\
$D_e$ (well depth)     & 0.37313     && 0.36725     & 0.36151     & 0.36402     & $E_\mathrm{h}$ \\
$D_0$ (with ZPE)       & 0.36796     && 0.36211     & 0.35641     & 0.35914     & $E_\mathrm{h}$ \\
$D_0$                  & 230.90      && 227.23      & 223.65      & 225.36      & kcal/mol \\
\tableline
\end{tabular}
\end{table}

To rigorously quantify this behavior, Table~\ref{n2spectro_combined} summarizes the spectroscopic constants extracted via a 10th-order Dunham polynomial expansion and a global Morse potential fit, tracking the performance from the modest cc-pVTZ up to the \stanoo{very} large cc-pV5Z basis set framework. The cross-methodological consistency between the local Dunham expansion and the global Morse fit \stanoo{shows that} the evaluated harmonic frequencies ($\omega_e$) and anharmonicities ($\omega_e x_e$) match within strict numerical tolerances.

Crucially, the comparison between the cc-pVTZ and cc-pV5Z results highlights the basis set convergence accelerated by the $\mathrm{F12Pi}$ operator. Going from the triple-zeta to the quintuple-zeta level shifts the Dunham harmonic frequency by a mere $15.9\,\mathrm{cm}^{-1}$ ($\omega_e = 2277.9\,\mathrm{cm}^{-1}$ vs. $2262.0\,\mathrm{cm}^{-1}$) and the net dissociation energy by 
$0.0058\,E_h$. After accounting for the natively derived zero-point energy (ZPE) corrections, the calculated net dissociation energies ($D_0 = 0.36796\,E_h$ for cc-pVTZ and $0.36211\,E_h$ for cc-pV5Z) are 
\stanoo{systematically approaching} 
the experimental spectroscopic limit of $0.35914\,E_h$. 
\stanoo{This smooth and systematic convergence profile} demonstrates that the $\mathrm{F12Pi}$ pseudo-excitations 
\stanoo{converge toward} the complete basis set (CBS) limit already within smaller basis set frameworks under a robust, collapse-free single-reference scheme.

\stanoo{However, an interesting structural anomaly is observed in Table \ref{n2spectro_combined}, where the CCSD0(T)-F12Pi framework yields a slightly elongated bond length $R_{e}$ and a lower harmonic frequency $\omega _{e}$, yet produces a larger dissociation energy $D_{e}$ compared to the experimental reference. This counterintuitive trend points to a known limitation of the current perturbative layer, mirroring the convergence behavior discussed later for methylene and cyclobutadiene. 
While the F12Pi pseudo-excitation operator successfully restores the short-range same-spin correlation at the doubles level, the non-iterative (T) triples correction is constructed from a cluster amplitude pool where the same-spin triplet components are strictly regularized to zero.
Because no complementary explicitly correlated terms are currently incorporated directly into the triples perturbation series, the longer-range components of the same-spin correlation channels remain insufficiently described in the asymptotic limit of triple-bond breaking. This leads to a minor underestimation of the correlation energy for the open-shell nitrogen atoms, artificially elevating the asymptotic baseline, \jozef{while increasing} $D_{e}$.
Presenting this imbalance as an inherent feature of the truncated triples configuration pool demonstrates that full quantitative synchronization across the potential landscape requires an analogous explicitly correlated F12 correction extended directly to the triples level.} 

\subsection{Symmetric Dissociation of $\mathrm{CO_2}$}
\begin{figure}[h]
\centering
\includegraphics[width=12cm]{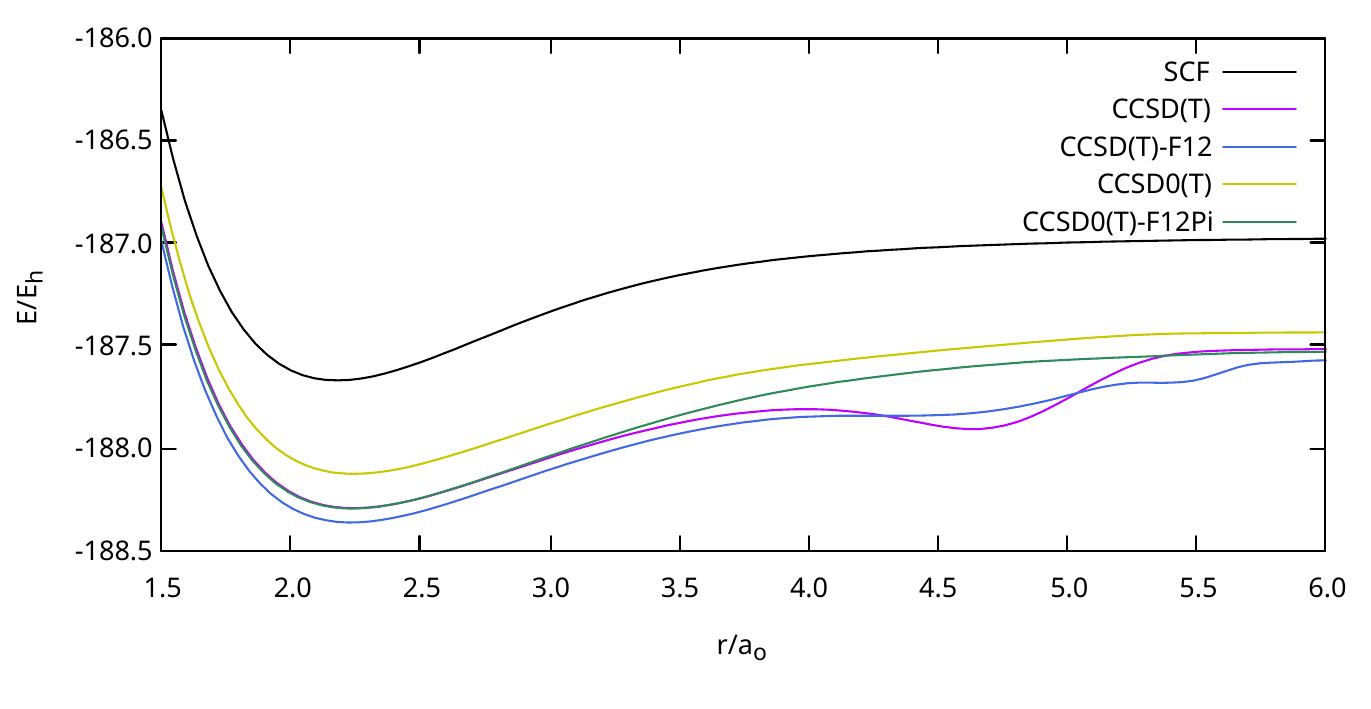}
\caption{\shota{Calculated potential energy curves for the $D_{\infty h}$-symmetric CO$_2$ with the cc-pVTZ basis set} as a function of bond length (in the atomic unit) calculated at the (black) RHF, (purple) CCSD(T), (blue) CCSD(T)-F12, (yellow) CCSD0(T) and (green) CCSD0(T)-F12Pi levels.}
\label{fig:CO2}
\end{figure}

The simultaneous breaking of four chemical bonds during the symmetric $D_{\infty h}$ stretching of $\mathrm{CO_2}$ imposes an even more severe stress test on the electronic wavefunction, characterized by strong multireference states and severe quasi-degeneracy. Figure \ref{fig:CO2} details the symmetric dissociation curves mapped at the RHF, CCSD(T), CCSD(T)-F12, CCSD0(T), and CCSD0(T)-F12Pi levels using the cc-pVTZ basis set. Standard single-reference CCSD(T) experiences a severe unphysical energy drop and an artificial hump spanning the $3.5\text{--}5.5\,a_0$ window. Supplementing this unstable reference with standard F12 pseudo-excitations (CCSD(T)-F12) fails to rectify the unphysical drop, instead introducing pronounced numerical roughness and oscillations across the intermediate potential surface.

\stanoo{ 
By regularizing the underlying cluster equations and simultaneously recovering the omitted same-spin triplet channels through the F12Pi pseudo-excitation operator, the parent wavefunction remains structurally uncorrupted even in regions of severe quasi-degeneracy. Because the underlying state-specific correlation tracks the physical domain correctly, the subsequent non-iterative perturbative triples correction remains globally stable and free of the typical single-reference instabilities, generating a physically sound and smooth trajectory without numerical artifacts.} 
The CCSD0(T)-F12Pi model exhibits 
quantitative agreement with CCSD(T) in the uncorrupted equilibrium region ($r \approx 2.2\,a_0$) while generating
a smooth, monotonic, and physically sound trajectory across the entire multireference domain.
In the asymptotic limit at larger separation distances ($r > 5.5\,a_0$), well beyond the quasi-degenerate multireference region, the curves for CCSD(T) and CCSD0(T)-F12Pi smoothly merge once again, 
with the conventional explicitly correlated CCSD(T)-F12 baseline remaining situated below them just as observed in the equilibrium region.
\stanoo{This smooth convergence underscores the qualitative consistency of our regularized framework as the system transitions back into a well-behaved single-reference regime.} 

\subsection{Multireference Benchmarks: Methylene and Cyclobutadiene}
To evaluate the performance of the method on systems characterized by profound intrinsic multiconfigurational character, calculations were extended to linear singlet methylene ($\mathrm{CH_2}$) and the transition state of cyclobutadiene ($\mathrm{C_4H_4}$) automerization, tracking basis set convergence up to the cc-pV5Z 
\stanoo{and cc-pVQZ levels, respectively.} 
The results are evaluated against state-specific Mukherjee multireference coupled-cluster (MkCCSD and MkCCSD(T)) benchmarks.

\begin{table}
\caption{Total energies in $E_\mathrm{h}$ of linear singlet methylene ($\mathrm{CH_2}$) \label{ch2table}}
\begin{center}
\footnotesize
\begin{tabular}{lccc|cc}
\tableline
Basis set  & CCSD0(T)-F12Pi & CCSD0(T)-F12Pi SP & CCSD0(T)    & MkCCSD(T)   & MkCCSD(T)-F12 SP \\
\cline{2-6}\nos
cc-pVDZ    & -38.995118 & -38.988092 & -38.952560 & -38.975627 & -39.010529      \\ 
cc-pVTZ    & -39.007371 & -39.005699 & -38.988403 & -39.019755 & -39.028987      \\ 
cc-pVQZ    & -39.011311 & -39.011441 & -38.997939 & -39.031149 & -39.034267      \\ 
cc-pV5Z    & -39.012672 & -39.013400 & -39.000826 & -39.034152 & -39.035622      \\ 
\tableline
Basis set  & CCSD0-F12Pi & CCSD0-F12Pi SP & CCSD0       & MkCCSD      & MkCCSD-F12 SP \\
\cline{2-6}\nos
cc-pVDZ    & -38.991582 & -38.984045 & -38.942493 & -38.974208 & -39.009270     \\ 
cc-pVTZ    & -39.001264 & -38.999177 & -38.978358 & -39.016898 & -39.026271     \\ 
cc-pVQZ    & -39.004382 & -39.004005 & -38.987903 & -39.027964 & -39.031157     \\ 
cc-pV5Z    & -39.005486 & -39.005622 & -38.990794 & -39.031037 & -39.032544     \\ 
\tableline
\end{tabular}
\end{center}
\end{table}

\stano{A rigorous analysis of the linear singlet methylene (\(\text{CH}_{2}\)) data in Table \ref{ch2table} reveals a remarkable contrast in the basis set convergence profiles between the pure doubles and the perturbative triples frameworks. For the explicitly correlated doubles, the CCSD0-F12Pi energy computed in the modest cc-pVDZ basis set (\(-38.991582\,E_\mathrm{h}\)) effectively lower-bounds the conventional \(\text{CCSD0}\) energy evaluated even within the \stanoo{very} large quintuple-zeta (cc-pV5Z) limit (\(-38.990794\,E_\mathrm{h}\)). This represents an extraordinary multi-zeta-level acceleration, showcasing how effectively the F12Pi pseudo-excitation operator saturates the interelectronic short-range correlation directly within a minimal orbital space.

In contrast, this massive acceleration threshold is noticeably attenuated when looking at the perturbative triples correction. Specifically, the CCSD0(T)-F12Pi energy in the cc-pVDZ basis (\(-38.995118\,E_\mathrm{h}\)) does not outpace the conventional \(\text{CCSD0(T)}\) quintuple-zeta results (\(-39.000826\,E_\mathrm{h}\)). This performance discrepancy stems from a fundamental theoretical limitation: the non-iterative (T) triples are perturbatively constructed from the underlying cluster amplitudes (Eq. \ref{eq:T3}). Because the parent \(\text{CCSD0}\) reference enforces strictly vanishing same-spin triplet biexcitation amplitudes to resolve the singlet-triplet instability, the configuration pool entering the triples perturbation series remains inherently incomplete. Crucially, while this truncation is seamlessly compensated at the doubles level by the F12Pi operator, no complementary explicitly correlated terms are incorporated into the triples correction framework. Consequently, the (T) correction lacks the necessary same-spin components to mirror the flat convergence profile observed for the doubles, highlighting an important physical trade-off in the regularized single-reference approach.}

\stano{It is worth noting in Table \ref{ch2table} that the energetic ordering between the fully relaxed and the standard SP-ansatz treatments reverses as the basis set is enlarged.}
\stanoo{For the series including the (T) correction, this reversal occurs between the triple- and quadruple-zeta levels, whereas for the pure doubles framework, the fully relaxed energy remains lower until the quintuple-zeta limit is reached. 
This behavior reflects the distinct physical foundations of the two treatments within finite orbital spaces. In smaller basis sets, the iterative optimization allows the F12 pseudo-excitation amplitudes to optimally relax 
and adapt to the incomplete computational space. In contrast, the fixed amplitudes utilized in the SP ansatz are inherently derived from the complete basis set (CBS) limit. As the underlying orbital basis approaches saturation, 
these asymptotic CBS-limit values become increasingly precise, allowing the SP framework to recover slightly more correlation energy compared to the fully relaxed variant under the Standard Approximation (SA), 
where the coupling terms involving the complementary auxiliary basis set space (CABS) are omitted rather than rigorously eliminated.} 

\begin{table}
\caption{Energies in $E_\mathrm{h}$ of cyclobutadiene \shota{at the} transition state \stanoo{geometry} \label{cyclobuttable}}

\begin{center}
\footnotesize
\begin{tabular}{lcc|ccc}
\tableline
Basis set  & CCSD0(T)-F12Pi SP & CCSD0(T)    && MkCCSD(T)   & MkCCSD(T)-F12 SP \\
\cline{2-6}\nos
cc-pVDZ    & -154.304513       & -154.096277 && -154.236142 & -154.378318      \\ 
cc-pVTZ    & -154.331890       & -154.210540 && -154.381172 & -154.418558      \\
cc-pVQZ    & -154.346025       & -154.244944 && -154.423188 & -154.435971      \\
\tableline
Basis set  & CCSD0-F12Pi SP & CCSD0       && MkCCSD      & MkCCSD-F12 SP \\
\cline{2-6}\nos
cc-pVDZ    & -154.282251    & -154.066148 && -154.212670 & -154.357236   \\
cc-pVTZ    & -154.301725    & -154.175546 && -154.346982 & -154.385628   \\
cc-pVQZ    & -154.313591    & -154.208900 && -154.386229 & -154.399579   \\
\tableline
\end{tabular}
\end{center}
\end{table}

\stano{In close analogy to the trends observed for singlet methylene, the cyclobutadiene automerization benchmarks compiled in Table 3 reinforce the remarkable basis-set convergence acceleration enabled by the explicit correlation. Evaluating the multireference transition state---a notoriously difficult stress-test for single-reference theories---the CCSD0-F12Pi method operating within the modest cc-pVDZ basis yields an energy of \(-154.282251\, E_\mathrm{h}\). This result successfully lower-bounds the conventional \(\text{CCSD0}\) energy computed in the much larger cc-pVQZ space (\(-154.208900\, E_\mathrm{h}\)). The fact that this substantial multi-zeta-level advantage holds true even for a highly degenerate system highlights the robustness of the F12Pi pseudo-excitation operator in saturating the dynamic components of the Coulomb hole without interfering with the underlying stability of the regularized model.

This hierarchical behavior similarly extends to the perturbative triples treatment, albeit under the same structural constraints discussed for singlet methylene. The \(\text{CCSD0(T)-F12Pi}\) energy evaluated within the modest cc-pVDZ basis (\(-154.304513\, E_\mathrm{h}\)) dramatically outpaces the conventional \(\text{CCSD0(T)}\) triple-zeta results and effectively lower-bounds the quadruple-zeta baseline (\(-154.244944\, E_\mathrm{h}\)). While the triples convergence profile remains less aggressively accelerated than that of the pure doubles - owing to the inherently truncated same-spin amplitude pool entering the (T) perturbation series -}
\stanoo{the overall energy profile systematically tracks the benchmark trajectories established by the vastly more expensive multireference MkCCSD and MkCCSD(T)-F12 frameworks. Although the single-reference energies lie above the multireference baseline beyond the double-zeta level, this systematic consistency firmly establishes the regularized single-reference formalism as an elegant and robust alternative for describing highly degenerate transition states.} 

\stano{This positioning underscores the inherent physical and mathematical trade-off of our approach. Because genuine multireference methods possess a vastly larger parameter space with multiple sets of cluster amplitudes, they naturally exhibit greater variational flexibility in strongly degenerate regions. While our single-reference framework cannot entirely match this extensive parameter-driven flexibility, 
the modified F12Pi operator successfully recovers the missing same-spin components, allowing the energies to approach the multireference baseline from above. This makes the present formalism an alternative that avoids the steep cost 
and active-space selection 
of full multireference implementations
by retaining the scaling of the parent framework.}

\subsection{
	{R}elated {CC} {M}odels}

\stano{Contrary} 
to the present approach, the 
\stanoo{formulation}
of CCSD0\cite{bulik} by 
Bulik \textit{et al.}
\stanoo{was originally introduced outside the context of the cluster-expansion analysis of closed-shell orbital theories\cite{hirao-closed} .}
\stano{In the same work, they also formulated an alternative model named CCSD1, where the cluster channels are spin-decoupled, selectively retaining the same-spin triplet pairing components while omitting the singlet counterparts}
\shota{
\begin{eqnarray}
	|\Psi_{\mathrm{CCSD1}}\rangle& =& \exp{(\hat{T}_1+\hat{T}^{[1]}_2)}|\Phi_0\rangle, \\
	\hat{T}^{[1]}_2& =& \frac{1}{2} \pi^{ab}_{ij}\hat{\mathbf{Q}}^{\dagger}_{ab}\hat{\mathbf{Q}}_{ij}, \\
	\hat{\mathbf{Q}}^{\dagger}_{ab}& =& (\hat{Q}_{ab}^{+\dagger}, \hat{Q}_{ab}^{0\dagger},\hat{Q}_{ab}^{-\dagger}), \\
	\hat{\mathbf{Q}}_{ij}& =& \begin{pmatrix} \hat{Q}_{ij}^+ \\ \hat{Q}_{ij}^0 \\ \hat{Q}_{ij}^- \end{pmatrix}, \\
		\hat{Q}_{ij}^+& =& a_{j\alpha}a_{i\alpha}, \\
	\hat{Q}_{ij}^0& =& \frac{1}{\sqrt{2}} (a_{j\alpha}a_{i\beta}-a_{i\alpha}a_{j\beta}), \\
	\hat{Q}_{ij}^- &=& a_{j\beta}a_{i\beta}.
\end{eqnarray}
}
\stano{They considered both CCSD0 and CCSD1 as regularized frameworks that truncate specific spin-excitation pathways. Gomez \textit{et al.} thus proposed recovering the omitted correlation components by re-coupling the spin-singlet and triplet-pairing channels:}
Optimizing the spin-triplet pair excitations referencing the CCSD0 ansatz (FSigCCSD) or the other way around (FPiCCSD),\cite{gomez2016-FSigCCSD}
\begin{eqnarray}
    |\Psi_{\mathrm{FSigCCSD}}\rangle = \exp{(\hat{T}^{[1]}_2)}|\Psi_{\mathrm{CCSD0}}\rangle, \\
    |\Psi_{\mathrm{FPiCCSD}}\rangle = \exp{(\hat{T}^{[0]}_2)}|\Psi_{\mathrm{CCSD1}}\rangle.
\end{eqnarray}
As Bintrim and Carter-Fenk evaluated pCCD, CCD0, FSigCCD as well as FPiCCD in 2025,\cite{bintrim_carter-fenk} these models in the "addition-by-subtraction" scheme are still considered as candidates for describing strong correlation and furthermore S$_1$/S$_0$ conical intersections. Therefore, the "addition-by-subtraction" electronic structure theories are worth evaluating by relating to the projected HF theories formulated by Hirao and Nakatsuji.\cite{hirao-closed} (Note that FSigCCD and FPiCCD were called CCDf0 and CCDf1, respectively, by Bintrim and Carter-Fenk.\cite{bintrim_carter-fenk})

Contrary to CCSD0, CCSD1 
\stanoo{cannot be structurally mapped to}
the projected HF theories, 
\stanoo{and a similar limitation applies to FPiCCSD.}
The situation of pCCSD is totally different. The pCCSD ansatz retains only the double excitations where electron pairs occupying the same orbital are excited to the same virtual orbital. Taking 
\shota{the Brueckner reference}
the pCCSD ansatz is expanded as
\begin{eqnarray}
    |\Phi_{\mathrm{pCCSD}}\rangle = C_0 |\Phi_{\mathrm{B}}\rangle + \sum_{ia} C_{ia}|\Phi_{ii}^{aa}\rangle + \cdots, \\
    |\Phi_{\mathrm{B}}\rangle = \det(\phi_1 \alpha \phi_1 \beta \cdots \phi_i \alpha \phi_i \beta \cdots \phi_n \alpha \phi_n \beta), \\
    |\Phi_{ii}^{aa}\rangle = \det(\phi_1 \alpha \phi_1 \beta \cdots \phi_a \alpha \phi_a \beta \cdots \phi_n \alpha \phi_n \beta).
\end{eqnarray}
As one sees, under the assumption that \shota{the Brueckner reference} can be approximated as the HF wavefunction, the pCCSD ansatz shares the 
paired
leading terms with those of EHF 
(\ref{eq:EHF})
and CMO
(\ref{eq:CMO}),
 the projected HF theories which overcome the real spin-triplet and imaginary spin-singlet stability and symmetry dilemmas, respectively. Therefore, pCCSD can be regarded as an approximation of the projected HF theories, which are orbital theories capable of describing strong correlation. Considering this correspondence and the scaling factor of pCCSD, $\mathcal{O}(N^3)$, with respect to the number of basis functions while that of CCSD0 is $\mathcal{O}(N^6)$, even though the 
pCCSD
ansatz lacks orbital-invariance, 
pCCSD-F12\cite{kedzuch2020}
may work for preliminary estimation before running calculations at the 
CCSD0-F12Pi
level.

\stano{The robust performance of the CCSD0-F12Pi methodology across the potential energy surface suggests its strong potential for describing excited states and $\mathrm{S_1/S_0}$ conical intersections. Furthermore, the combination of conventional cluster expansions by $\hat{T}$ and pseudo-excitations by $\hat{R}$ in our ansatz may enable CCSD0-F12Pi to overcome the breakdown near $\mathrm{S_1/S_0}$ conical intersections inherent to normal CC theories, as explained by Kjønstad and Koch based on the vanishing-component theorem.}\cite{phase_effect}

Therefore, we are currently working on extending the 
CCSD0-F12Pi
theory to excited states in the equation-of-motion coupled-cluster (EOM-CC) scheme.\cite{stanton_bartlett_eom-cc,Bartlett2007,bartlett_review2012} In practical calculations in the future, especially excited-state dynamics simulations based on 
"EOM-CCSD0-F12Pi",
the $\mathcal{O}(N^6)$ scaling will be a bottleneck. Nevertheless, we do not think that it is a drawback because "EOM-pCCSD-F12", whose scaling is $\mathcal{O}(N^3)$, may work for preliminary estimation.

It should be emphasized that, in cases of open-shell systems, CCSD0\cite{gomez2016-ccsd0open} does not correspond to the projected HF theories\cite{hirao-open} and thus  
\stanoo{the configuration analysis above does not apply.}
For applying CCSD0-F12Pi to open-shell systems, the IP-EOM-CC\cite{stanton_ip-eom-cc1999} and EA-EOM-CC\cite{musial_ea-eom-cc2003} schemes seem promising.

\section{Conclusion}

\stanoo{We recapitulated the cluster-expansion analysis of closed-shell orbital theories by Hirao and Nakatsuji, and employed their classification of the HF instabilities to identify which correlation channel is responsible for the breakdown of single-reference coupled-cluster theory in stretched geometries. 
By comparing these orbital frameworks with singlet-paired coupled-cluster singles and doubles, we showed that the capability of CCSD0 to describe strong electron correlation 
is consistent with
its sharing of leading paired configurations with the projected Hartree–Fock theory.} 
\stanoo{This correspondence is structural rather than energetic: CCSD0 is projective and size-consistent whereas CMO is variational and size-inconsistent, and CCSD0 lowers the energy relative to RHF irrespective of whether the RHF solution is stable.} 

\stanoo{We supplemented the electron correlation omitted in CCSD0 by employing the F12 theory. Among the short-range correlation recovered by F12, 
we focus on the same-spin component that is structurally absent in CCSD0. This targeted mathematical treatment utilizes the F12 framework to efficiently retrieve the specific triplet correlation components 
that were regularized out from the parent single-reference ansatz, while simultaneously maintaining the intrinsic convergence acceleration toward the complete basis set limit.} 

Rigorous tests on the homonuclear triple-bond cleavage of $\mathrm{N_2}$ and the multiple-bond symmetric dissociation of $\mathrm{CO_2}$ visually and numerically confirm that our framework entirely eliminates the unphysical artifacts, humps, and catastrophic energy collapses exhibited by standard CCSD(T) and CCSD(T)-F12 methodologies. 
\stano{Furthermore, applications to the highly degenerate transition state of cyclobutadiene and the linear singlet methylene benchmark demonstrate that CCSD0-F12Pi significantly accelerates the basis-set convergence profiles, yielding an effective multi-zeta level improvement over the conventional parent references while retaining a strict, computationally elegant single-reference scaling.}
\stanoo{Moreover, the systematic approach of the CCSD0(T)-F12Pi dissociation energies toward the spectroscopic data as the basis set approaches the quintuple-zeta limit reflects the sound physical behavior of the present formalism}
for electronic structure calculations for strongly-correlated systems.

\stano{For a more accurate treatment in CCSD0-F12Pi, implementing with auxiliary bases approximation or three electron integrals might be useful if there were reasonable algorithms available with favorable computational scaling. 
Additionally, further systematic improvements could be achieved by extending the triexcitations correction framework to incorporate the missing same-spin triplet components, which are currently omitted in both the parent CCSD0(T) and CCSD0(T)-F12Pi approaches, by introducing an analogous explicitly correlated F12 correction directly at the triples level.}

\stanoo{Taken together, these results indicate that a regularized single-reference ansatz, combined with an explicitly correlated correction for the omitted same-spin channel, provides a practical route to describing strongly correlated closed-shell systems—one that requires no active-space selection and retains the $\mathcal{O}(N^6)$ scaling of conventional CCSD. Extensions to explicitly correlated FSigCCSD and to the equation-of-motion framework for excited and open-shell states are currently in progress.}

\section*{Conflicts of interest}
There are no conflicts to declare.

\section*{Author contributions}
\textbf{Stanislav Ked\v{z}uch}: Conceptualization (equal); Data curation (equal); Formal analysis (equal); Investigation (equal); Methodology (lead); Software (equal); Validation (equal); Visualization (equal); Writing -- original draft (equal); Writing -- review \& editing (equal).
\textbf{Shota Tsuru}: Conceptualization (equal); Formal analysis (supporting); Funding acquisition (equal); Investigation (equal); Methodology (equal); Project administration (lead); Resources (supporting); Validation (supporting); Visualization (equal); Writing -- original draft (equal), Writing -- review \& editing (supporting).
\textbf{Takahito Nakajima}: Conceptualization (equal); Funding acquisition (equal); Resources (lead); Supervision (equal); Writing -- review \& editing (lead).
\textbf{Jozef Noga}: Conceptualization (equal); Formal analysis (lead); Methodology (equal); Software (equal); Supervision (equal); Validation (equal); Writing -- review \& editing (equal).

\shota{
\section*{Acknowledgements}
This work was supported by JSPS KAKENHI (Grant Nos. JP25K00248 and JP25K03396). S.T. thanks S. Coriani at the Technical University of Denmark, A. Sunaga and J. Hasegawa at Hokkaido University as well as T. Sato at the University of Tokyo for fruitful discussions. \jozef{J.N. appreciates support from the Slovak R\&D Agency under project No. APVV-25-0785.}
}

\bibliography{references}

\end{document}